\documentclass[twocolumn]{aastex62}

\newcommand\s{$\sim$}
\newcommand{\e}{\'{e}}

\usepackage{xcolor}

\shorttitle{Major mergers are not the dominant trigger for high-accretion AGNs at $z\sim2$}
\shortauthors{Marian et al.}

\begin{document}

\title{Major mergers are not the dominant trigger for high-accretion AGNs at $z\sim2$}

\correspondingauthor{Victor Marian}
\email{marian@mpia.de}

\author[0000-0003-1733-9281]{Victor Marian}
\affiliation{Max-Planck-Institut f\"ur Astronomie,
K\"onigstuhl 17,
69117 Heidelberg, Germany}
\affiliation{International Max Planck Research School for Astronomy \& Cosmic Physics at the University of Heidelberg}

\author[0000-0003-3804-2137]{Knud Jahnke}
\affiliation{Max-Planck-Institut f\"ur Astronomie,
K\"onigstuhl 17,
69117 Heidelberg, Germany}

\author[0000-0001-6462-6190]{Mira Mechtley}
\affiliation{School of Earth and Space Exploration, Arizona State University, P.O. Box 871404, Tempe, AZ 85287-1404, USA}

\author[0000-0003-3329-1337]{Seth Cohen}
\affiliation{School of Earth and Space Exploration, Arizona State University, P.O. Box 871404, Tempe, AZ 85287-1404, USA}

\author[0000-0003-2901-6842]{Bernd Husemann}
\affiliation{Max-Planck-Institut f\"ur Astronomie,
K\"onigstuhl 17,
69117 Heidelberg, Germany}

\author[0000-0003-4665-8521]{Victoria Jones}
\affiliation{School of Earth and Space Exploration, Arizona State University, P.O. Box 871404, Tempe, AZ 85287-1404, USA}

\author[0000-0002-6610-2048]{Anton Koekemoer}
\affiliation{Space Telescope Science Institute, 3700 San Martin Drive, Baltimore, MD 21218, USA}

\author[0000-0002-6660-6131]{Andreas Schulze}
\affiliation{National Astronomical Observatory of Japan, Mitaka, Tokyo 181-8588, Japan}

\author[0000-0002-5027-0135]{Arjen van der Wel}
\affiliation{Sterrenkundig Observatorium, Universiteit Gent, Krijgslaan 281 S9, B-9000 Gent, Belgium}
\affiliation{Max-Planck-Institut f\"ur Astronomie, 
K\"onigstuhl 17, 
69117 Heidelberg, Germany}

\author[0000-0002-8956-6654]{Carolin Villforth}
\affiliation{University of Bath, Department of Physics, Claverton Down, BA2 7AY, Bath, United Kingdom}

\author[0000-0001-8156-6281]{Rogier A. Windhorst}
\affiliation{School of Earth and Space Exploration, Arizona State University, P.O. Box 871404, Tempe, AZ 85287-1404, USA}

\begin{abstract}

Research over the past decade has shown diminishing empirical evidence for major galaxy
mergers being a dominating or even important mechanism for the growth of
supermassive black holes in galaxies and the triggering of optically or X-ray selected active galactic nuclei (AGN).

We here for the first time test whether such a connection exists at least in
the most `plausible' part of parameter space for this mechanism: the highest specific
accretion rate broad-line AGNs at the peak epoch of black hole activity around $z=2$.
To that end we examine 21 galaxies hosting a high accreting black hole ($L/L_\mathrm{edd} > 0.7$) 
observed with HST/WFC3 and 92 stellar mass- and redshift-matched
inactive galaxies taken from the CANDELS survey.

We removed the AGN point sources from their host galaxies and avoided bias in visual classification by adding and then subtracting mock point sources to and from the comparison galaxies, producing matched residual structures for both sets. The resulting samples were joined, randomized, and subsequently visually ranked with respect to
perceived strength of structural distortions by 10 experts. The ensuing individual rankings were combined
into a consensus sequence and from this we derived merger fractions for
both samples. With the merger fractions $f_\mathrm{m,agn}$ = 0.24 $\pm$ 0.09 for the AGN host galaxy sample and $f_\mathrm{m, ina}$ = 0.19 $\pm$ 0.04 for the inactive galaxies, we find no significant
difference between the AGN host galaxies and inactive galaxies. 
Also, both samples display comparable fractions of disk-dominated
galaxies. These findings are consistent with previous studies for different
AGN populations, and we conclude that even black hole growth at the highest specific
accretion rates and at the peak of cosmic AGN activity is not predominantly caused by major
mergers.

Considering AGN and merger timescales we conclude that it is highly unlikely that a potential time lag could significantly affect our results. We also calculate the inverse view and infer that only $\sim$20\% of all merging galaxies have high-accretion AGN activity during the ongoing merger. Eventually, we also find no significant connection between post-starburst galaxies and AGNs. 

\end{abstract}


\section{Introduction} \label{sec:intro}

Active galactic nuclei (AGN) and their feedback could play an important role in the evolutionary sequence of galaxies
by truncating or even triggering and supporting star formation, therefore severely affecting the evolution of a galaxy \citep[e.g.][]{98_silk_agn_feedback, 12_fabian_agn_feedback_review, 18_harrison_agn_feedback_20_years_review}. 
A tight physical coupling between central black hole (BH) and hosting galaxy environment is currently only observed for a number of galaxy clusters in terms of suppressive feedback in the form of growing bubbles in the intracluster medium driven by radio jets, which can hinder the cooling of gas and subsequently star formation \citep{10_blanton_agn_feedback_gal_cluster}. While a general physical "coevolution" between BH and galaxy is not required to explain the tight scaling relations between masses of the two \citep[e.g.][]{
03_marconi_bh_bulge_mass,
04_häring_scaling_bh_mass_host_mass,
09_jahnke_bh_bulge_total_mass_evo, 
10_bennert_scaling_bh_mass_host_lum_z_2, 
11_bennert_scaling_bh_mass_host_mass_z_2,
11_jahnke_origin_bh_scal_rel_central_theorem, 
12_beifiori_bh_vel_disp_rel,
13_graham_bh_mass_lum_relation,
13_mcconnell_bh_mass_gal_prop_scal_rel}, the vast amounts of energy being emitted during phases of nuclear activity can in principle lead to a number of feedback effects, which depend for instance on the outflow sizes and geometries, if the expelled gas leads to increased or reduced star formation, or if the gas is swept out of the galaxy or just prevents halo gas from cooling \citep[][and references therein]{18_husemann_agn_feedback}.

Considering these observations and the fact that every massive galaxy hosts a supermassive black hole in its center \citep{13_kormendy_coevolution_smbh_host_review}, it is of utmost interest to understand the processes involved in the growth of black holes. While the immediate environment and feeding of a black hole is described by the well accepted unified AGN model \citep{93_antonucci_unified_agn_model, 95_urry_unified_agn_model, 15_netzer_revisiting_unified_model}, the mechanisms transporting gas from kilo-parsec scales to the central parsecs of a galaxy, to be potentially accreted by the BH are less well understood.  

For decades, it was a widely accepted scenario that the occurrence of AGNs should be a consequence of a merger of gas-rich galaxies of comparable mass, i.e.\ a major merger \citep{92_barnes_interacting_galaxies_review, 96_sanders_lirgs_review}. As a result of the tidal interactions and associated gravitational torques the gas would most probably lose its angular momentum and therefore a subsequent infall into the central region. This process is first extensively described in the seminal work of \cite{88_sanders_merger_agn_paradigm} connecting ultraluminous infrared galaxies (ULIRGs) and AGNs. Since then this paradigm was frequently investigated and allegedly corroborated throughout the last three decades via simulations and models \citep{05_springel_model_feedback_bh_merger, 06_hopkins_merger_driven_model, 08_hopkins_cosmo_framework_coevo, 08_somerville_merger_agn_model, mcalpine_18_arxiv_eagle_major_merger_dominant_low_z, 18_weigel_merger_agn_model} as well as more recent observations and studies of specific local AGN populations  
\citep{10_koss_merging_agn_swift_sample, 13_cotini_merger_agn_conn_observed_cas_local, 13_sabater_effect_interaction_nuclear_activity, 15_hong_corr_merger_agn} and especially high luminosity AGNs at different redshifts \citep{08_urrutia_merger_agn_obscured_z_1, 12_schawinski_obscured_z_2_discs_not_major_merger, 12_treister_high_lum_agn_merger_con, 15_glikman_merger_most_lum_conn_z_2, 16_fan_most_lum_qso_high_merger_z_3, 18_donley_merger_driven_lum_high_z_obscured_agn, 18_goulding_high_lum_agn_conn_time_variability}.

However, over the last years the relevance of major interactions and the connection to the growth of BHs has been extensively tested in a systematic way, leading to the contrary finding in which the merging of two gas-rich galaxies is playing a sub-dominant part for the occurrence of an AGN. Neither for the bulk of X-ray detected and optically observed AGNs across cosmic time 
\citep{09_gabor_AGN_host_morphos_z_0.3_1, 
09_georgakakis_AGN_disc_gal_not_mergers, 
11_cisternas_BH_Growth_z_1} 
nor for AGNs at low or intermediate luminosities 
\citep[$L_\mathrm{X}\leq10^{43}$erg s$^{-1} $;][]{05_grogin_agn_z_0.4_1.3_no_merger_conn, 
11_allevato_x_ray_bl_agn_intermediate_lum_no_merger,
11_schawinski_AGN_z_2_SMBH_growth_disk_gal,
12_kocevski_merger_connection_z=2_x_ray,
13_boehm_agn_hosts_z_0.7_no_merger,
15_cheung_agn_bar_connection_mod_lum,
15_cisternas_agn_bar_connection, 
15_rosario_host_gal_x_ray_agn_z_2.5_herschel,
17_goulding_agn_bar_connection}
 an observational connection in form of a triggering and fueling by major mergers could be found.
Studies examining AGNs with high luminosities 
\citep[$L_\mathrm{X}\geq10^{43}$erg s$^{-1} $;][]{14_karouzos_high_ir_lum_agn_no_merger_conn, 
14_villforth_no_merger_agn_conn_z_0.7,
17_villforth_host_gal_lum_z_0.6_no_merger_con,
17_hewlett_z_evo_merger_agn},
 BHs possessing the highest masses \citep{16_mechtley_agn_merger_most_massive} or heavily obscured AGNs \citep{12_schawinski_obscured_z_2_discs_not_major_merger} came to the same conclusion. Recently also for AGNs considered to be in their earliest evolutionary stage no dominant connection to major mergers could be detected \citep{18_villforth_felobal_qso_no_mergers}.

What almost all of these publications have in common is a matched control sample of inactive galaxies to compare the AGN sample with.
Only by taking such a matched comparison sample into account, a statement regarding a possible  enhancement of the merger fraction for AGN host galaxies over the normal incidence for inactive galaxies is valid. 

To further examine this observed independence of BH growth on major mergers in more detail, we analyze in this work one of the central parts of parameter space that has so far remained untested: AGNs with the highest Eddington ratios, which is equivalent to the highest specific accretion rates, i.e.\ AGNs with the highest accretion rates relative to the respective black hole masses.
We examine this particular subpopulation of AGNs at a redshift $z\sim2$, which corresponds to the epoch of peak AGN activity \citep{00_boyle_peak_agn_activity, 15_aird_accr_rate_density_peak_z_2} and cosmic star formation \citep{14_madau_cosmic_SF_history}.

\begin{figure*}[t]
\centering
\includegraphics[width = 18cm]{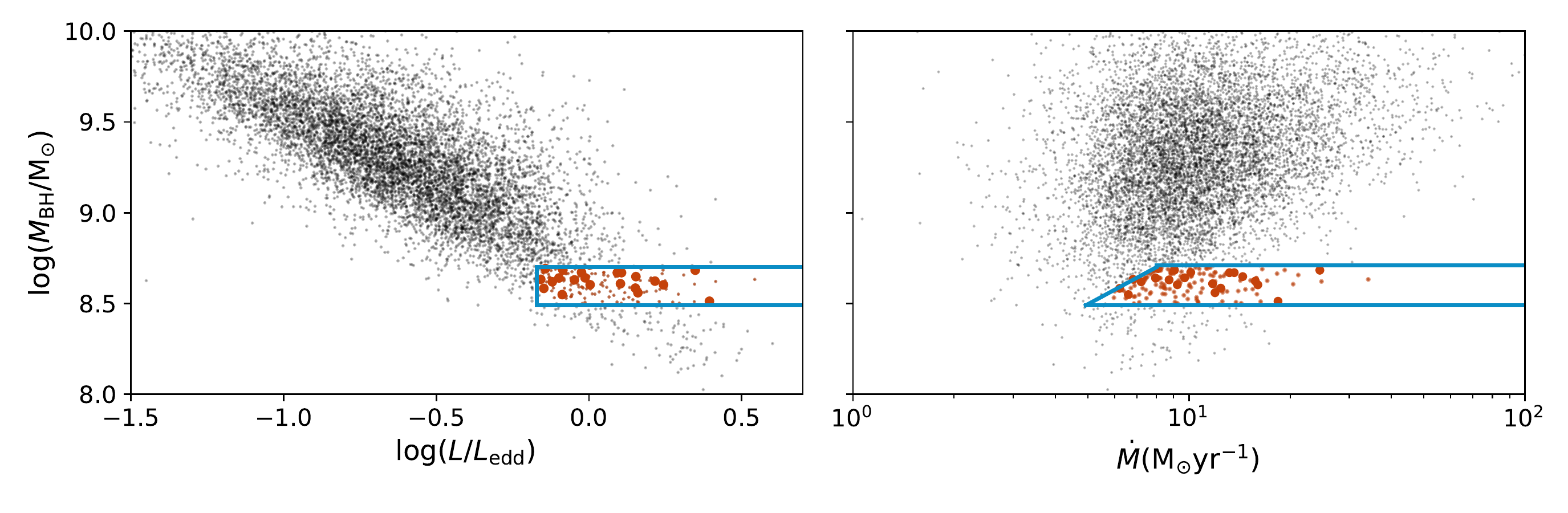}
\caption{\textit{Left}: Black hole mass vs. Eddington ratio for the complete sample of AGNs at redshift $1.8\leq z\leq2.2$ (black dots). Overplotted are our limits in mass and Eddington ratio (blue box) for the parent sample (small red dots) and our final selection of AGNs (large red dots). \textit{Right}: Black hole mass vs. mass accretion rate per year for the same samples indicating that our final selection consists of AGNs possessing the highest specific accretion rates.}
\label{fig:QSO_sample}
\end{figure*}

It sounds intuitive and plausible that at least black holes belonging to this particular subset need amounts of gas that can only be delivered by gravitational interaction of two gas-rich galaxies of equal mass. In order to test this conjecture, we compare 21 AGN host galaxies to a matched sample of 92 inactive comparison galaxies at this particular redshift. Since major merger features for the CANDELS sample are visible out to $z\sim2.5$ \citep[][and references therein]{13_kaviraj_visibility_merger_features_candels} and our observations of the AGN host galaxies provide the same depth and S/N, we can test the existence of a potential causal sequence between major mergers and the growth of black holes. We quantify these signatures of major mergers like tidal torques and tails or asymmetries as a proxy to determine if our source is or recently was part of a gravitational interaction. 
We combine both samples after accounting for the AGN nuclei in the images, have experts rank the joint sample by distortion strength, and then split the sample again into AGN and non-AGN to determine separate fractions of distorted galaxies.

We also turn our attention to two questions, which arise when considering differences in the merger and AGN timescales. First, if there is the possibility of a time lag between merger features and an AGN, which would lead eventually to observing the latter, while signs of the gravitational encounter are not detectable anymore. Secondly, how it would impact our results when we consider the opposite in the form that merger features are clearly visible, but an AGN connected to this process is currently not, as proposed in \citet{18_goulding_high_lum_agn_conn_time_variability}. Finally, we also examine a potential connection between post-starburst galaxies and AGNs, since such galaxies show significant signs of a recent merger, but also need a mechanism to stop the recently increased star formation. 

The paper is structured as follows. In Sect. 2 we describe the data and modeling of our two samples. In Sect. 3 we give our analysis, while in Sect. 4 we compare our results with respect to similar studies, and discuss various implications like the average morphologies of our two samples, potential issues arising when considering different timescales for an AGN and a major merger or if post-starburst galaxies are more prone to host an AGN.
Finally, Sect. 5 summarizes our findings. We adopt cosmological parameters $\Omega_{\Lambda} = 0.7, \Omega_0 = 0.3$ and $h = 0.7$. All magnitudes are given in the AB system \citep{83_oke_ab_system}. 

\section{Data} \label{sec:data}

\subsection{AGN host galaxies} \label{subsec:AGN_sample}

We construct our initial parent sample by selecting targets from a subset of the SDSS DR7 Quasar catalog \citep{09_abazajian_sdss_dr7, 10_schneider_sdss_qso_z, 11_shen_SDSS7_qso_spec_props}, 
which were uniformly selected by the target algorithm described in \citet{02_richards_SDSS_qso_selection} \citep[see also][]{06_richards_SDSS_qso_target_algorithm}, and constrain it to a mass range of $8.5\leq \text{log}(M_\mathrm{BH}/\mathrm{M_{\odot}})\leq 8.7$ by using the broad-line Mg\ \textrm{II} measurements of the virial black hole masses by \citet{11_shen_SDSS7_qso_spec_props}. Furthermore, we select only AGN with an Eddington ratio $(L/L_\mathrm{edd}) \geq$ 70\% and in a redshift range of  $1.8\leq z\leq2.2$. Moving to even higher redshifts at $z \gtrsim2$ would impede reliable BH mass determination due to a lack of available AGN SDSS spectra with sufficient quality of Mg~\textrm{II} line measurements.
Combined with the fact of reaching the limits in terms of resolution and PSF stability, this is therefore likely the highest redshift range at which high accretion AGNs can be reliably investigated for galaxy properties with the Hubble Space Telescope (HST). 
With these constraints we avoid the lower mass limit of SDSS, which lies at $\textrm{log}(M_\mathrm{BH}/\mathrm{M_{\odot}}) \sim 8.3$ \citep{08_shen_biases_bh_masses_sdss, 08_vestergaard_lower_mass_limit}, but still probe the AGN population, which is providing at least half of all BH mass growth in the early universe and exhibits the highest Eddington ratios, i.e.\ the highest specific accretion rates (Fig.~\ref{fig:QSO_sample}, left panel). For the calculation of the mass accretion rates $\dot{M}$ (Fig.~\ref{fig:QSO_sample}, right panel) we use the bolometric luminosities presented in the SDSS DR7 Quasar catalog and  assume an efficiency parameter $\eta = 0.1$. 

We designed the experiment so that sample sizes would permit to significantly differentiate between a $\geq$50\% merger fraction for the AGN sample and an (assumed) 20\% fraction for inactive galaxies. For this goal and \s90 comparison galaxies (see below), 21 AGN were sufficient to detect a difference at a $>$99.7\% probability level, i.e.\ $>$3$\sigma$ confidence. 
Hence, we reduce the resulting initial sample of 147 targets, which satisfy our conditions to 21, randomly selected AGNs to avoid any bias selection. These limits in black hole mass and Eddington ratio/mass accretion rate are shown as blue boxes in Fig.~\ref{fig:QSO_sample}, the initial 147 AGNs and finally 21 selected AGNs, are shown as small and large red dots, respectively.
A visual inspection of the spectra of those targets reassures the robustness of the black hole mass determinations.

\begin{deluxetable*}{lhccccccDc}
\tablenum{1}
\tablecaption{AGN sample properties\label{tab:qso_prop}}
\tablewidth{0pt}
\tablehead{
\multicolumn2l{Target AGN} & \colhead{Redshift} &  \colhead{M$_\mathrm{I}$} & \colhead{$L_\mathrm{bol}$} & \colhead{$L_{3000 }$} & \colhead{FWHM} & \colhead{$M_{\mathrm{BH}}$} & \multicolumn2c{log($L/L_{\mathrm{edd}}$)} & \colhead{$\dot{M}_\mathrm{acc}$}  \\
\multicolumn2l{(SDSS J)} & \colhead{$z$} & \colhead{mag} & \colhead{$\textrm{ergs}^{-1}$} & \colhead{log($\mathrm{L_{\odot}}$)} & \colhead{Mg~\textrm{II} (km s$^{-1}$)} & \colhead{log($\mathrm{M_{\odot}}$)} & \nocolhead{} & \nocolhead{} & \colhead{$\mathrm{M_{\odot} yr^{-1}}$} 
}
\decimalcolnumbers
\startdata
083253.96+434712.5 & & 1.841 & -26.51 & 46.58 & 12.28 & 2100 & 8.5 & -0.09 & 6.6  \\ 
084632.93+272343.2 & & 2.039 & -27.71 & 47.15 & 12.65 & 1900 & 8.7 & 0.35 & 24.6  \\ 
091555.62+294741.6 & & 2.086 & -27.14 & 46.76 & 12.47 & 2100 & 8.7 & -0.02 & 10.2  \\ 
101415.24+404113.5 & & 1.960 & -27.21 & 46.92 & 12.56 & 2000 & 8.6 & 0.15 & 14.3  \\ 
102549.18+220435.6 & & 2.064 & -27.10 & 46.88 & 12.54 & 2000 & 8.7 & 0.09 & 13.1  \\ 
110621.43+104432.5 & & 1.860 & -26.85 & 46.72 & 12.43 & 2000 & 8.6 & 0.00 & 9.2  \\ 
114415.49+452445.2 & & 1.879 & -26.48 & 46.61 & 12.32 & 2200 & 8.6 & -0.12 & 7.2  \\ 
123623.34+281158.3 & & 1.953 & -27.00 & 46.89 & 12.55 & 2000 & 8.7 & 0.11 & 13.7  \\ 
130418.72+240935.1 & & 1.894 & -27.45 & 46.96 & 12.67 & 1700 & 8.6 & 0.24 & 15.9  \\ 
130913.27+381751.0 & & 1.807 & -26.61 & 46.66 & 12.36 & 2200 & 8.6 & -0.10 & 7.9  \\ 
141449.07+421649.1 & & 1.846 & -26.63 & 46.59 & 12.29 & 2300 & 8.6 & -0.16 & 6.8  \\ 
141551.86+110316.3 & & 1.873 & -27.18 & 46.85 & 12.56 & 1800 & 8.6 & 0.15 & 12.4  \\ 
141811.67+271103.2 & & 2.023 & -27.05 & 46.55 & 12.43 & 2000 & 8.6 & -0.15 & 6.2  \\ 
142632.62+295948.8 & & 1.934 & -26.80 & 46.74 & 12.39 & 2200 & 8.6 & -0.01 & 9.8  \\ 
142723.96+533858.7 & & 1.960 & -26.84 & 46.83 & 12.42 & 1900 & 8.6 & 0.16 & 12.0  \\ 
151312.67+303542.7 & & 1.943 & -26.69 & 46.71 & 12.31 & 2400 & 8.7 & -0.09 & 9.0  \\ 
153014.46+070430.9 & & 1.966 & -26.57 & 46.66 & 12.30 & 2500 & 8.7 & -0.14 & 8.1  \\ 
153946.63+271105.7 & & 2.145 & -27.23 & 46.83 & 12.45 & 2000 & 8.6 & 0.10 & 11.7  \\ 
154632.34+223637.8 & & 1.984 & -27.04 & 46.95 & 12.66 & 1800 & 8.6 & 0.22 & 15.9  \\ 
155631.49+080051.7 & & 1.914 & -27.25 & 47.02 & 12.64 & 1600 & 8.5 & 0.39 & 18.2  \\ 
162411.38+220428.0 & & 2.004 & -26.72 & 46.70 & 12.37 & 2200 & 8.6 & -0.05 & 8.7  \\ 
\enddata
\tablecomments{Properties of the AGNs in our sample. Columns 1-6 are taken from the catalog of \citet{11_shen_SDSS7_qso_spec_props}. Column 1: SDSS object designation (J2000.0); Column 2: redshift determined in the SDSS DR7 \citep{10_schneider_sdss_qso_z}; Column 3: k-corrected \citep{06_richards_SDSS_qso_target_algorithm}, absolute I-band magnitude, normalized at $z=2$; Column 4: bolometric luminosity; Column 5: luminosity at 3000 \AA; Column 6: FWHM of Mg \textrm{II} in the broad component; Column 7: black hole virial masses using Mg \textrm{II}; Column 8: Eddington ratio, calculated using the bolometric luminosity $L_{bol}$ of Column 4 and the black hole masses $M_{\mathrm{BH}}$ of Column 7; Column 9: Mass accretion rate in solar masses per year.}
\end{deluxetable*}

Subsequently, these 21 sources were observed with HST/WFC3 in the F160W band, corresponding to $V$ in rest-frame (HST program ID 14262, PI K. Jahnke). The exposure time for each target observation was one orbit (\s 40min), divided into a 6$\times$400s dithering pattern. 
With a resulting lower magnitude limit of 24 mag, we are thoroughly capable to detect features of major mergers, like tidal tails or asymmetries at $z\sim2$ \citep{10_van_dokkum_f160w_mag_limit_merger, 12_ferreras_f160w_mag_limit_merger}.  

\begin{figure*}
\centering
\includegraphics[width = 18cm]{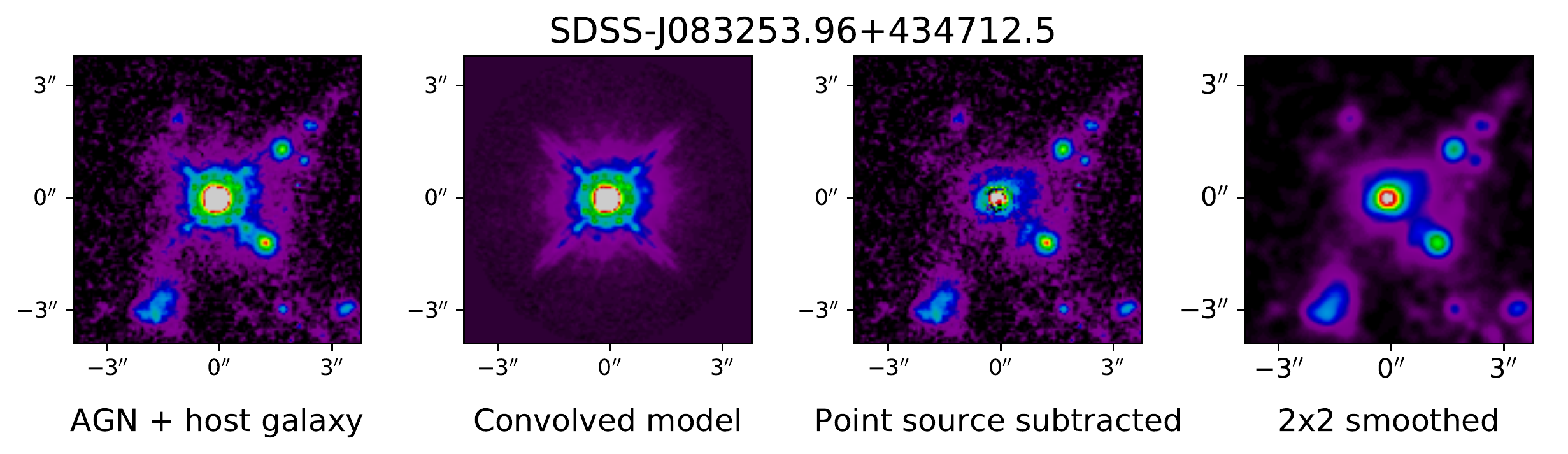}
\caption{Different steps of our AGN/host galaxy modeling process shown for one sample source: In the first column we show the original, processed HST/WFC3 image, the second column displays the PSF convolved point-source model and the third and fourth column show the host galaxy after point-source subtraction with the original sampling and smoothed by a $\sigma$=2px Gaussian, respectively. All images are displayed with the same arcsinh scaling. The remaining 20 AGNs of the sample are shown in appendix~\ref{appendix:qso_models_figs}.}
\label{fig:qso_showcase}
\end{figure*}

In Table~\ref{tab:qso_prop} we summarize the properties of our AGNs. Aside from the SDSS source designation we state the corresponding redshifts $z$, as well as the k-corrected \citep{06_richards_SDSS_qso_target_algorithm} absolute I-band magnitudes, normalized at $z=2$, the bolometric luminosity $L_\mathrm{bol}$, the rest-frame luminosities at 3000\AA\ $L_{3000}$, the FWHM of the single-epoch measurements of Mg~\textrm{II} in km\,s$^{-1}$, and the respective black hole masses $M_\mathrm{BH}$, all taken form the catalog presented in \citet{11_shen_SDSS7_qso_spec_props}. Furthermore, we present the calculated corresponding Eddington ratios $L/L_\mathrm{edd}$ and mass accretion rates $\dot{M}_\mathrm{acc}$.

\subsection{Inactive galaxies} \label{subsec:Ina_sample}

To construct our comparison sample of inactive galaxies, we make use of the exhaustive observations done within the CANDELS Multi Cycle Treasury program \citep{11_grogin_candels, 11_koekemoer_candels_overview}. To ensure the best possible match to our AGN sample, we select galaxies with the same properties as our AGN host galaxies within the GOODS-S, GOODS-N, AEGIS, COSMOS and UDS fields, choosing sources in the same redshift and mass range, and observed with the same instrument and filter. To that end, we apply the same redshift cuts and use HST/F160W observations, which posses a comparable S/N to our single orbit AGN observations. To estimate the needed stellar masses of our AGN host galaxies, we utilize the $M_\mathrm{BH}$--$M_*$ scaling relation \citep{04_häring_scaling_bh_mass_host_mass} and include a normalization factor of 2 for a redshift evolution out to $z=2$ \citep{10_bennert_scaling_bh_mass_host_lum_z_2, 11_bennert_scaling_bh_mass_host_mass_z_2, 11_schulze_bh_bulge_relation_selection_effects}. Subsequently, we randomly select out of a few thousand \citep{14_van_der_wel_3D-HST+CANDELS_mass_size_distr_since_z_3} potential candidates 92 galaxies with $9.2 < \text{log}(M_*/\mathrm{M_{\odot}}) < 11.7$ to comprise our sample of inactive galaxies, with the individual mass estimates taken from the catalog presented by the 3D-HST survey \citep{12_brammer_3dhst, 14_skelton_3dhst_phot_cat, 16_momcheva_3dhst_spectra_etc}. We chose this rather wide range in mass to account for the uncertainties in the black hole mass determinations, which in turn have an impact on the host mass estimates. Considering these uncertainties we predict an identical range in stellar mass for our AGN host galaxies as we have chosen for our comparison sample. 
However, to determine any influence of stellar mass on our results, we repeat our analysis with a mass constrained subsample ($10.4 < \text{log}(M_*/\mathrm{M_{\odot}}) < 11.5$) of inactive galaxies.

\subsection{Data reduction and Bayesian 2D modeling}

Since we are interested in detecting potential merger features of our AGN host galaxies, it is mandatory to subtract the nuclear AGN light component that is superposed on the stellar light from the host galaxy, which can be brighter by a factor of 5-40 than the underlying host galaxy \citep{04_jahnke_qso_host_sf_from_color, 04_jahnke_uv_young_stars_in_qso_host_high_z,  04_sanchez_agn_color, 07_kotilainen_bright_nucleus_factor, 08_schramm_bright_nucleus_factor, 09_jahnke_bh_bulge_total_mass_evo, 14_mechtley_phd}. Hence, a good PSF model is essential to achieve a satisfactory subtraction of the central AGN component.
Generally, the best choice is to build a PSF model from stars within the same image and with the same colour as the AGN. However, since this is not possible for all our exposures, and because WFC3 suffers from optical imperfections, such as coma or astigmatism, we construct additional PSF models from archival data of high S/N stars \citep[see][]{16_mechtley_agn_merger_most_massive}. We make sure that these observations match our AGN observations in resolution, readout-mode, and dithering pattern. Our final set consists of 16 different PSF models.

As a next step, the individual exposures of each object of our two samples and selected PSFs were combined using software within \texttt{drizzlepac} \citep[following approaches outlined in ][]{02_koekemoer_drizzle, 12_gonzaga_drizzlepac_handbook} resulting in a 0\farcs060 pixel scale, corresponding to $\sim$~0.5~kpc at $z\sim2$. We create variance maps for all our images as well as for the PSF models, which are mandatory for the subsequent modeling. Excluding variance maps from our modeling process would lead to an underestimation of errors, since the flux contribution of the central point-source is significantly higher than the sky. 

We create 2D two-component models of our objects utilizing the same PSFMC algorithm\footnote{see \url{https://github.com/mmechtley/psfMC} for additional information, examples and documentation} \citep{14_mechtley_phd} as described and used in \cite{16_mechtley_agn_merger_most_massive}. We use a point-source model and a S\e rsic profile to simultaneously fit the central AGN component and underlying host galaxy respectively. Although the host galaxies may not be perfectly described by a S\e rsic profile, it is a sufficient approximation, as we can determine potential asymmetries or tidal features. Furthermore, the advantage of using this two component approach is twofold: First we assure that the contributing flux of the central source is not overestimated and avoid an eventual PSF over-subtraction. Secondly, we also keep the computational time needed for the fits within reasonable limits.

Contrary to most of the tools adopted to fit galaxies in 2D, our algorithm allows the user to input a prior probability distribution for the fitted model parameters of the different components, e.g. position in x and y, total magnitude, S\e rsic index, position angle, etc. As a result, posterior probability distributions of all the fitted parameters with the corresponding covariances are produced. Additionally, the algorithm does not assume an error-free PSF, but implements the respective PSF variance maps during the convolution process. To that end, we deploy the HST focus model \citep{11_cox_hst_focus_model} to determine for each source the time-dependent telescope focus and find the best matching PSF within our PSF library. 

For the actual fit we mask and exclude neighboring galaxies, which central flux peaks do not coincide with the central source in the respective images. The resulting point-source component is then subtracted from the HST/WFC3 image to remove the AGN, and just display the host galaxy beneath. In Fig.~\ref{fig:qso_showcase} we show from left to right for one of our 21 AGN sources the original, drizzled HST/WFC3 image, the convolved point-source model and the point-source subtracted model showing the host galaxy, once with the original sampling and once smoothed by $\sigma$=2px Gaussian. All four images are shown in the same arcsinh scaling. Comparable plots for the remaining 20 AGNs are shown in appendix~\ref{appendix:qso_models_figs}. 

To achieve our goal for galaxies in both samples exhibiting the same nuclear residuals, we add a synthetic point source on top of the flux's center position of the inactive galaxies by randomly choosing one of our PSFs. If needed, we add additional noise to match the observed AGNs and then perform the same fitting process as for the AGN sample. 

As a final result we now have two samples not only matched in mass, redshift, S/N, instrument and filter, but also displaying the same central point-source subtraction residuals.

\section{Morphological analysis \& merger fractions} \label{sec:analysis}

After the modeling process our final joint sample comprises of 113 images of AGN hosts and inactive galaxies. As a next step to derive the merger fractions we have asked experts, who have extensive experience in working with imaging data of galaxies\footnote{The ranking has been done by the coauthors Cohen, Husemann, Jahnke, Jones, Koekemoer, Marian, Schulze, van der Wel, Villforth and Windhorst} to visually analyze the images and identify possible features reflecting major mergers, and subsequently rank the galaxies from most to least distorted. Such an approach to visually detect features from gravitational interactions has been successfully applied by us and others \citep{11_cisternas_BH_Growth_z_1, 12_kocevski_merger_connection_z=2_x_ray, 16_mechtley_agn_merger_most_massive}, and is still superior to any automated algorithm \citep{09_jogee_visual_inspection_superior}. Each expert will have their own subjective interpretation of what a major or minor disturbance might look like, but the "blind" analysis forces the same criteria for both, mixed, samples. Hence, personal biases towards the appearance of mergers do not play a significant role in our analysis. To remove further systematic biases, every classifier has been provided his/her own randomized sample order and naming. Additionally we request them to flag the images, which in their opinion show a non-resolved source, due to the image being noise dominated or displaying only artifacts from the point source subtraction. To that end we provided images of eight point source subtracted stars and also included four of them in our joint sample of galaxies to assess the `quality' of the individual rankings. While only one galaxy has been marked by five experts as non resolved with all remaining galaxies receiving less flags, the point source subtracted star images have each been flagged five to nine times. In total out of 1170 individual rankings, 165 are non-resolved flags (with two experts accounting for 96 of them), with 35 flags belonging to the five aforementioned objects (one galaxy and four stars). For the subsequent creation of the consensus ranking, we excluded these objects, which left us with 112 galaxies.

To determine whether and how strong the combination method might impact the results, we apply three different techniques to combine the individual rankings. For our first approach we follow \citet{16_mechtley_agn_merger_most_massive} and weigh the individual ranks of each galaxy. As our second and third method we adopt the Borda count \citep{13_emerson_borda_count} and the Schulze method \citep{11_schulze_schulze_method, 18_schulze_schulze_method}, respectively (see appendix~\ref{appendix:details_methods} for  details on the different combination methods and on how we implement them). 

To derive the merger fractions, we split our three combined rankings at a particular cut-off rank, below which a visual inspection does not show any obvious presence of gravitational interaction features. To be consistent, we chose the same cut off rank for every one of our consensus sequences at rank 22. With a separation at this cutoff-rank into a set of galaxies showing obvious merger features and a sample free of such features, we obtain for all three consensus rankings identical numbers. Consequently, we verify that neither the choice of combination method nor of our distinction between merging and non-merging sources have any significant impact on our final results. A detailed analysis can be found in appendix~\ref{appendix:fraction_dependences}.

We obtain identical results for all three consensus rankings, which provide the following numbers:
\begin{itemize}
\renewcommand\labelitemi{--}
\item 5 AGN host galaxies show merger features
\item 16 AGN host galaxies show no such features
\item 17 inactive galaxies show merger features
\item 74 inactive galaxies show no such features
\end{itemize}

It should be noted that the order of objects does not have to be and is indeed not the same between the three different combined lists. Still, the objects below our chosen cut-off rank at position 22 vary only slightly: All of the 5 AGNs and 11 out of the 17 comparison galaxies are identical for the three consensus rankings.

The probability distributions of the derived merger fractions can be described by a beta distribution. Since $a$, the number of galaxies exhibiting merger features, and $b$, the number of not distorted galaxies are integers, the probability distribution function of the beta distribution can be written as

\begin{equation}
f(x) = \frac{(a+b+1)!}{a!\,b!} x^{a-1} (1-x)^{b-1}.
\end{equation}

\begin{figure}[t]
\includegraphics[width = 9cm]{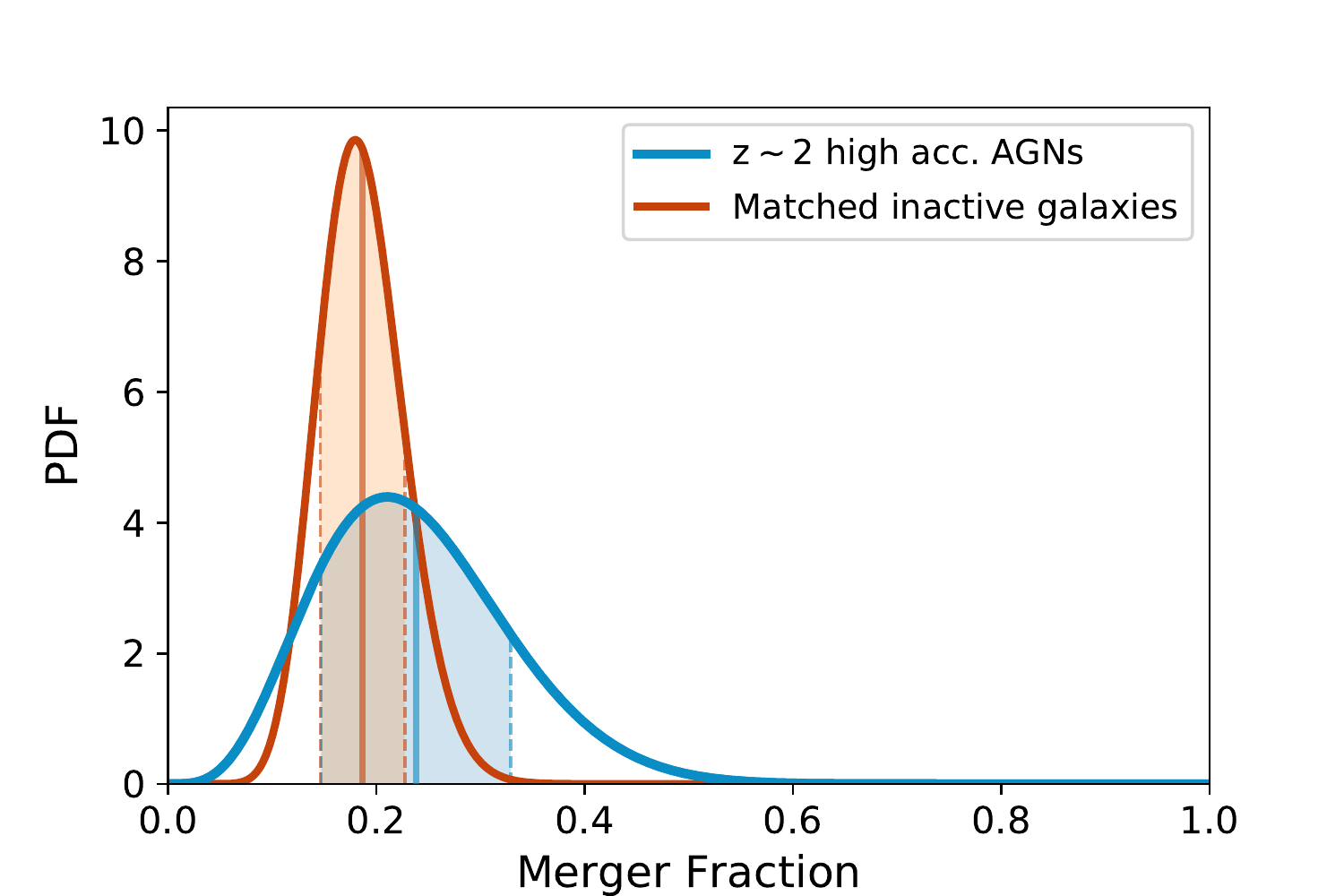}
\caption{Probability distributions for the inferred merger fractions of our $z\sim2$, high-accretion AGN host galaxies (blue) and inactive galaxies (red). The solid line show the means of the respective merger fractions, while the dashed lines denote the central 68\% confidence intervals. \label{fig:merger fractions}}
\end{figure}

The probability distributions for the inferred merger fractions are shown in Fig.~\ref{fig:merger fractions} in red for the sample of inactive galaxies and in blue for the AGN host galaxies. The solid lines display the expected values, i.e.\ means for both samples, the dashed lines and shaded regions the associated 68\% confidence intervals. The most probable merger fractions, i.e.\ the means of the distributions, which are described by $a/(a+b)$, are $f_\mathrm{m,agn}$ = 0.24 $\pm$ 0.09 for the AGN host galaxy sample and $f_\mathrm{m, ina} =\ $0.19 $\pm$ 0.04 for the inactive galaxies. 
This difference is indistinguishable within 1$\sigma$. A Welch’s t-test testing the hypothesis that both samples have identical means yields a p-value of $p \sim 0.12$.

\section{Discussion} \label{sec:discussion}

We calculated the merger fractions for a sample of 21 AGNs and 92 inactive, matched comparison galaxies at $z\sim2$. Since our retrieved merger fractions are insensitive to the combination method or cut-off rank, we consider pure Poisson noise as the dominating source of uncertainty. With that assumption we find that for our AGN sample the absolute merger fraction is equal or below 50\% with $\sim99.4\%$ confidence. 

In addition, we tested if the merger fraction depends on Eddington ratio. To that end we split our AGN sample into two subsets with the cut made at $L/L_\mathrm{edd}=1$, which -- except for one source -- corresponds to a separation at $\dot{M}=10\mathrm{M}_{\odot}\mathrm{yr}^{-1}$. Coincidentally, this bisects our sample, with ten AGNs having an Eddington ratio $L/L_\mathrm{edd}<1$. If major mergers are indeed responsible for the transport of large gas masses to the center and therefore supporting such high accretion rates, we would expect that all or most of our five AGN host galaxies classified as mergers also belong to the subset showing accretion rates $\dot{M}>10\mathrm{M}_{\odot}\mathrm{yr}^{-1}$. However, only two of the five host galaxies are above this threshold, whereas even the five highest accreting AGNs are classified as non-disturbed.  

As it is also very likely that we identified some non-distorted sources as mergers and vice versa, we do not interpret the derived fractions as absolute values. However, since this misclassification-bias applies to both samples equally, we can examine the relative differences without any constraints. With the merger fractions derived in Sect.~\ref{sec:analysis} we immediately obtain at first glance a slight enhancement with respect to the AGN sample of $f_\mathrm{m,agn}/f_\mathrm{m,ina} = $ 1.26 $\pm$ 0.43. However, this fraction is not significantly different from unity, and in combination with our finding that the AGN merger fraction is with high probability well below 50\% and the lack of a trend of merger features with accretion rate, this leads us to the conclusion that major mergers are not predominantly responsible for the triggering and growth of black holes with the highest specific accretion rates at $z\sim2$. This result supports previous findings of major mergers playing a sub-dominant role in triggering AGN activity, but includes now even black holes with the highest specific accretion rates.

To rule out any impact on the specific selection of stellar mass, we repeat our calculations with a mass-constrained subsample of our comparison galaxies with $10.4 < \text{log}(M_*/\mathrm{M_{\odot}}) < 11.5$. Such a mass range corresponds to the range in stellar masses of our AGN host galaxies, if we would not consider any uncertainties in black hole mass estimates, when applying the scaling relation presented in Sect.~\ref{subsec:Ina_sample}. This approach reduces our number of inactive galaxies by 18, neglecting 14 galaxies with lower masses and four with higher masses. By applying the same cuts with respect to the number of flags (see Sect.~\ref{sec:analysis}) and an identical cut-off rank at 22, the merger fractions for our AGN sample is unchanged, whereas the fraction for our comparison galaxies increases to $f_\mathrm{m,ina} = 0.23 \pm 0.05$. This value can now be assumed to be identical with the one derived for the AGN sample and could be explained twofold: Since 14 of the 18 neglected galaxies are at the lower mass end, their surface brightnesses and S/N are correspondingly lower, which makes it more difficult to detect merger features. Additionally, due to their apparent smaller sizes, they are also more affected by the point-source subtraction, which only amplifies the chance of misclassification. 

However, since we ignore these galaxies, we also change the overall rankings, which in turn means we should actually quantify how many of these sources were within the original merger sample, i.e.\ below a cut-off rank of 22. For the averaging approach and the Schulze method the number of galaxies amounts to 4, while only 2 are considered merging when considering the consensus ranking compiled via Borda counts. If we now adapt the overall cut-off rank correspondingly, i.e.\ lower it by 3 to be at rank 19, we retrieve a merger fraction for the inactive galaxies of $f_\mathrm{m,ina} = 0.19 \pm 0.05$. This number matches our initial result for the distortion fraction of our comparison sample, and we interpret this behavior as an insignificant dependence of merger fractions on stellar mass. 

\vspace{1cm}

\subsection{Comparison with previous studies}

We find that our results are consistent with recent studies at lower Eddington ratios, higher BH masses, or lower redshifts, and support the emerging picture of a general lack of a causal connection between severe gravitational interactions of gas-rich galaxies and occurrences of AGNs.  

For their sample of AGNs dominating the BH growth since $z\sim1$ \citet{11_cisternas_BH_Growth_z_1} found a similar distortion fraction. Approximately 85\% of their 140 X-ray selected and with HST/ACS observed type 1 and 2 AGNs at $0.3<z<1.0$ display no distinct merger features. Comparable values are found by \citet{09_gabor_AGN_host_morphos_z_0.3_1}, as well as \citet{12_kocevski_merger_connection_z=2_x_ray} for 72 intermediate luminosity AGNs ($L_\mathrm{X} \sim 10^{42-44}$ erg s$^{-1}$) at $1.5<z<2.5$ by using infrared data from HST/WFC3. Through visual classification they find a merger fraction for their AGN hosts of $16.7^{+5.3}_{-3.5}$\%, which is consistent with the fraction derived for their comparison sample of 216 inactive galaxies. Without quantifying any specific distortion fractions, \citet{11_allevato_x_ray_bl_agn_intermediate_lum_no_merger} and \citet{15_rosario_host_gal_x_ray_agn_z_2.5_herschel} infer similar results for low- and moderate-luminosity AGNs as do \citet{05_grogin_agn_z_0.4_1.3_no_merger_conn} and \citet{13_boehm_agn_hosts_z_0.7_no_merger} by using various proxies, like asymmetry, concentration, the Gini coefficient or the $M_{20}$ index to assess the morphologies of the AGN host galaxies.
\citet{17_hewlett_z_evo_merger_agn} and \citet{17_villforth_host_gal_lum_z_0.6_no_merger_con} found supporting evidence that major mergers are not strongly connected to AGN at high luminosities ($L_\mathrm{X}\sim 10^{43-45}$ erg~s$^{-1}$) either.
Both studies report for their X-ray selected and HST observed samples consistent merger fractions of \s 20\% for both, AGN hosts and inactive galaxies. 

Considering more specific AGN populations \citet{12_schawinski_obscured_z_2_discs_not_major_merger} analyzed 28 heavily dust-obscured galaxies suspected to host AGNs at $z\sim2$ with HST/WFC3. They describe a similar merger fraction between 4\% and at most 11--25\% for their sample of AGN host galaxies, but do not directly compare this to an inactive galaxy comparison sample. Another study by \citet{18_donley_merger_driven_lum_high_z_obscured_agn} found a higher fraction of merger morphologies among more luminous IR-dominated, heavily obscured, AGN, but confirmed that mergers are not a dominant source of fuelling in their sample of less obscured X-ray AGN. Lastly, \citet{16_mechtley_agn_merger_most_massive} targeted AGNs with high BH masses ($M_\mathrm{BH} = 10^9 - 10^{10} \mathrm{M_{\odot}}$) at $z\sim2$. In that paper, our analysis of 19 HST/WFC3 observed AGN and comparison to 84 matched inactive galaxies, did not reveal a significant enhancement in merger fractions of $f_\mathrm{m,agn}/f_\mathrm{m,ina} = 1.30 \pm 0.33$, which is identical to our result. 

Although we only consider the relative difference of the merger fractions and advise caution interpreting those number as absolute values, the result for $f_\mathrm{m, ina}$ is in excellent agreement with our expectations. Based on previous studies, we anticipated the general merger fraction of galaxies at $z\sim1$ to be about a factor 2 higher compared to $z=0$ \citep[e.g.][]{04_lin_major_merger_rates_z_1.2, 10_bridge_major_merger_rates_z_1.2}. Going to larger redshifts simulations predict a further increase, which is proportional to $(1+z)^2$ \citep{08_fakhouri_merger_rates_millenium_sim}, resulting in an estimated merger fraction of $\sim$20\% for inactive galaxies.
Using cosmological hydrodynamic simulations, \citet{18_steinborn_sim_merger_significant_or_not} obtain comparable merger fractions with less than 20\% of AGN host galaxies displaying any features of a gravitational encounter with another galaxy. \citet{18_snyder_illustris_merger_evolution} retrieved similar values by using synthetic images from the Illustris simulation and comparing it to CANDELS observations. Both datasets returned consistent results at $z\sim2$. Finally, 
\citet{17_ventou_muse_merger_fraction_evo} report comparable values using spectroscopic MUSE (Multi Unit Spectroscopic Explorer) observations in the Hubble Ultra Deep Field and Hubble Deep Field-South by applying close pair counts. For their subsample of massive galaxies $(\text{log}(M_{*}/\mathrm{M_{\odot}}) > 9.5)$, which is almost identical to the mass range of our samples they find merger fractions of 0.232$^{{+0.112}}_{{-0.056}}$ and 0.195$^{{+0.142}}_{{-0.081}}$ for the redshift bins $1\leq z\leq2$ and $2\leq z\leq4$, respectively. So, although our goal was not to retrieve absolute values for the merger fractions, but rather a relative difference, our estimated merger fractions are well consistent with these other recent observations.

In summary, our results are consistent with a set of previous studies examining a large part of parameter space within the AGN population. The consensus findings suggest a scenario where there is no obvious connection between major mergers and the triggering and growth of black holes. This results appears to be independent of AGN properties like luminosity, mass or Eddington ratio.

\subsection{Morphological sample comparison} \label{Discussion:morph_comp}

Since both our samples are chosen randomly, we assume that the morphologies of our inactive and host galaxies are equally distributed. To test this hypothesis and to determine the fraction of disk-dominated host galaxies, we calculate the average S\e rsic index over all individual S\e rsic profile fits. For our two samples we retrieve almost identical values for the mean S\e rsic indices, with our AGN host galaxies and inactive galaxies yielding $n_\mathrm{agn} = 5.42 \pm 3.46$ and $n_\mathrm{ina} = 5.24 \pm 3.2$, respectively. Assuming that the galaxies are only either disk- or bulge-dominated, one can use the Beta distribution (see Sect.~\ref{sec:analysis}) again to infer the corresponding fractions.  We make the distinction of disk- and bulge-dominated galaxies at a S\e rsic index of $n = 2.5$ \citep{03_shen_sersic_index_distinction}, and retrieve fractions of disk-dominated sources of $f_\mathrm{disk,agn} = 0.29 \pm 0.10$ and $f_\mathrm{disk, ina} = 0.22 \pm 0.04$ for the AGN sample and the comparison sample respectively. While these absolute assessments should be treated with caution, since we only used a single S\e risc profile to describe the individual galaxies, we assume that the relative difference of the fraction between the AGN sample and our control sample is representative, as both samples are fitted by using the same approach. Although the ratio of the total number counts of our two samples are almost 1:5, an ensuing two-sample Kolmogorov-Smirnov (KS) test with a resulting $p = 0.62$ supports our assumption that both samples stem from the same parent distribution.

The statistically identical fractions of disk-dominated galaxies are a further argument against the merger scenario, since we would expect a lower fraction of disk-like AGN host galaxies due to the merging process. Our findings are consistent with the results of \citet{09_georgakakis_AGN_disc_gal_not_mergers}, who found that $30 \pm 9$ per cent of AGN hosts are dominated by a disk-like structure compared to the total AGN space density, and $23 \pm 6$ per cent of the total luminosity density at $z\sim1$. We cannot reproduce the high percentage (\s 80\%) of disk-like host galaxies that \citet{11_schawinski_AGN_z_2_SMBH_growth_disk_gal} describe for their intermediate luminosity AGN sample at $z\sim2$. Nevertheless, by performing a two-sample KS test they found no difference between the distributions of S\e rsic indices for their AGN sample and their matching comparison sample either. 

Since it is a primary proxy for galaxy mass we also examined the magnitudes of the S\e rsic components returned by the modeling process. Although we want to caution once again that we included this component primarily to avoid an over-subtraction by the point source we find on average $m_\mathrm{S\text{\'{e}}rsic} = 21.4 \pm 0.6~\mathrm{mag}$ and $m_\mathrm{S\text{\'{e}}rsic}~=~21.7 \pm 0.6~\mathrm{mag}$ for the AGN sample and inactive galaxies, respectively. This very good agreement between the two samples confirms that our two samples are indeed well matched in stellar mass.

\subsection{Undetectable merger features: Potential time lag and isophote analyses}\label{subsec:time_lag}

One often discussed caveat is a potential time lag between a major merger process and the putative onset of AGN activity. A discrepancy of that form would wash out any merger features before we could detect the AGN, meaning that the AGN was indeed triggered by the gravitational encounter, but obscured while the visible features of a merger were still visible. 
However, considering recent observations and simulations we can argue against this point. 
The usual lifetimes of AGNs are described to be $10^6-10^8$ yr \citep{04_martini_qso_lifetime, 05_hopkins_qso_lifetime, 07_shen_qso_lifetime, 09_hopkins_qso_lifetime, 13_conroy_qso_lifetime, 15_cen_qso_lifetime} with 50--75\% -- correlating inversely with X-ray luminosity -- of this timespan being spent in an obscured state \citep{07_gilli_obscured_unobscured_AGN_fraction}, while recent observations show that the visibility of the features of a major merger are of the order of $10^8 - 10^9$ yr \citep{06_conselice_merger_timescale, 08_lotz_merger_timescale, 14_ji_merger_timescale, 18_solanes_merger_timescale}. 
Presuming that the active phase starts at the same time as the merger, i.e.\ at first passage, we conclude that for the longest living AGNs with maximum obscuration and a minimum lifetime for the merger features there is an overlap of at least a couple of $10^7$ yr in which the AGN phase and galaxy distortions are simultaneously visible. We expect this conservative minimum estimate to be also valid for low host galaxy masses with the shortest merger timescales. For our AGN sample with higher host galaxy masses it is more probable that the merger features are visible several 100 Myr longer than the AGN phase itself.  
Additionally, instead of observing comparable fractions of disk-dominated galaxies in both samples, we would expect an increase in bulge-dominated galaxies for our AGN host galaxies sample, since they would have undergone a higher number of major mergers.

As an additional test to detect potential post-merger features, we performed an elliptical isophote analysis \citep{2013_astropy_I, 2018_astropy_II, photutils} of our AGN host galaxies as well as of our point source subtracted inactive galaxies and compared the resulting centroid positions to the corresponding point source positions returned by the modeling process. To that end we calculate the average radial distances between the modeled point source coordinates and the central positions determined by the outer isophotes for both samples, i.e.\ $\langle r \rangle_\mathrm{dist, agn}$ for the AGN host galaxies and $\langle r \rangle_\mathrm{dist, ina}$ for our comparison sample and test whether $\langle r \rangle_\mathrm{dist, agn}$ is within the errors of $\langle r \rangle_\mathrm{dist, ina}$. 
In the case of an enhanced fraction of major merger events for our AGN sample, we would expect $\langle r \rangle_\mathrm{dist, agn}$  to be larger than $\langle r \rangle_\mathrm{dist, ina}$ due to possible merger induced offsets between AGN positions and actual flux centers of the underlying host galaxies. For a robust measurement of the central positions we fit our isophote ellipses out to \s 15kpc and visually check if the central x and y coordinates remain constant within the last measurements. In all cases the coordinates converge to their final value at \s 5kpc. We also apply a sigma clipping to account for eventual nearby sources.  

\begin{figure}[t]
\includegraphics[width = 9cm]{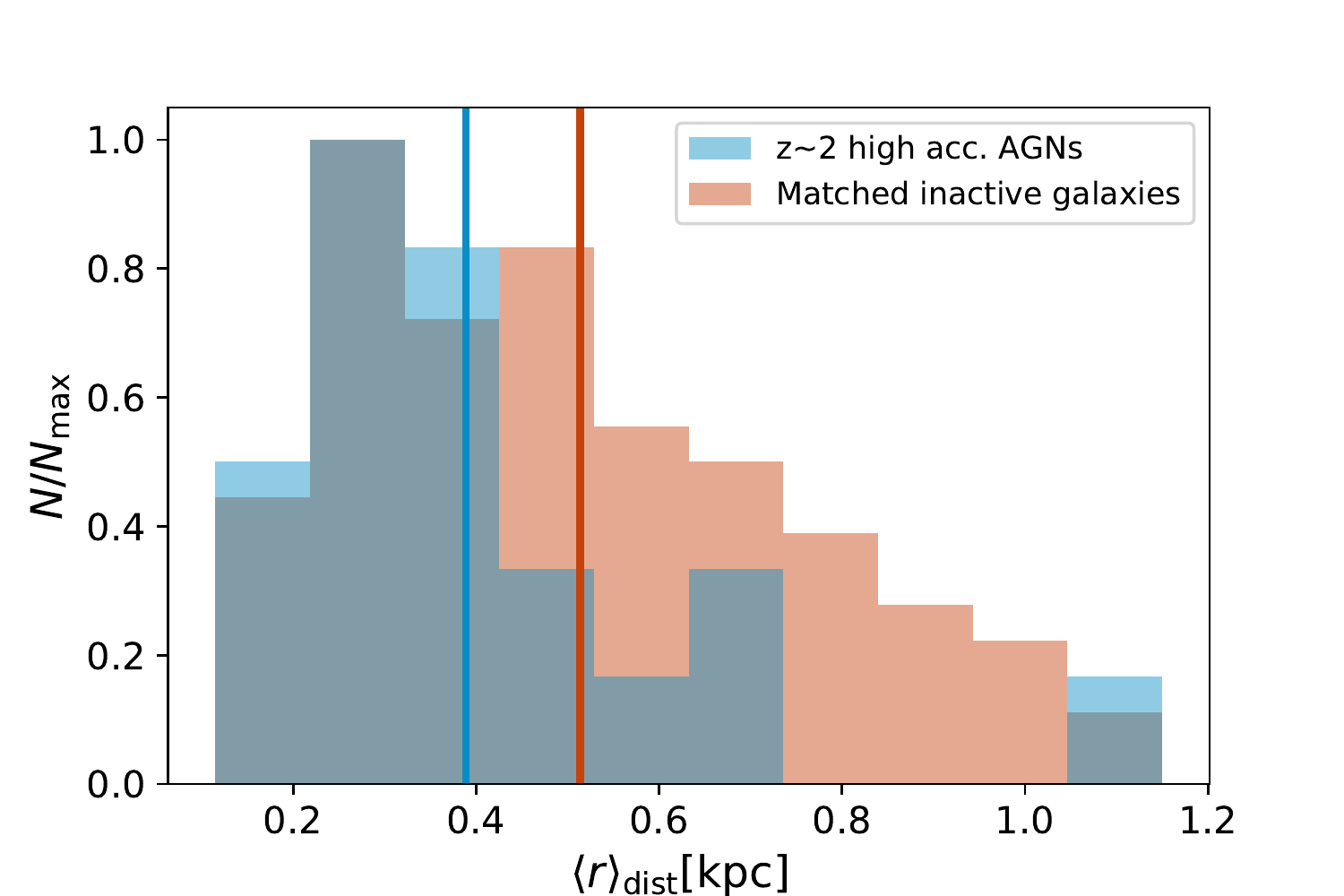}
\caption{Distributions of the radial distances in kpc between the centroid positions of the isophote ellipses and the point source coordinates from the model for our AGN sample in blue and for our comparison sample of inactive galaxies in red. Both histograms are normalized to $N/N_\mathrm{max}$. The solid lines in the corresponding colors denote the mean radial distances $\langle r \rangle_\mathrm{dist, agn}$ and $\langle r \rangle_\mathrm{dist, ina}$.
\label{fig:rad_distance_distributions}}
\end{figure}

For our AGN sample we retrieve an average radial distance of $\langle r \rangle_\mathrm{dist, agn} = 0.4 \pm 0.2$kpc, while our analysis for the comparison sample yields $\langle r \rangle_\mathrm{dist, ina} = 0.5 \pm 0.2$kpc. In Fig.~\ref{fig:rad_distance_distributions} we show the corresponding distributions of radial distances in kpc $\langle r \rangle_\mathrm{dist, agn}$  in blue and $\langle r \rangle_\mathrm{dist, ina}$ in red for our AGN and comparison sample, respectively. To facilitate a direct comparison, both histograms are normalized to their maximum amplitude and the solid lines present the aforementioned average radial distances. 

Considering that for the inactive galaxies the synthetic point sources were added on the flux weighted centers of each galaxy, we would expect in a perfect case an $\langle r \rangle_\mathrm{dist, ina} = 0$. However, the isophotal analyses were performed on images after point source subtraction -- a process which introduced inevitable residual artifacts. The distance distribution for the inactive galaxies hence represents the error for this measurement. We conclude that the results we retrieve for the AGN sample lies within the error introduced by the point source subtraction, since $\langle r \rangle_\mathrm{dist, agn} < \langle r \rangle_\mathrm{dist, ina}$. Therefore, this test also yields within our resolution limit no enhancement of merger incidents for our AGN sample.

\subsection{Intermittence of AGN activity}\label{subsec:intermittence}

Contrary to the scenario above where the AGN lifetime appears longer compared to the observability of merger features,  we can consider the other extreme case where, although there has been a merger induced AGN phase with distinct merger features visible, no AGN is detectable -- due to intermittance of accretion.
Such an approach would solve the inconsistent findings concerning a potential connection between major mergers and AGNs and is presented in \citet{18_goulding_high_lum_agn_conn_time_variability}. They explain the different results presented in previous studies as listed in Sect.\ref{sec:intro} by those findings depending on the stage of the merger and a directly-resulting AGN variability. During close passages of the galaxies, the arising torques are sufficient to feed the AGN, but when the separation increases again, the AGN activity dies down as well. As a result, the tidal features are still visible, but the central BH is not sufficiently fueled to be active, and only gets reignited after ensuing passes. This implies that in this scenario some of the sources we classified as merging, but inactive galaxies, actually host an active, but intermittent AGN. That is, the BH currently appears or is indeed inactive, but had already a phase of high accretion, and will achieve such high accretion rates again in the future, which would place it then well within our AGN sample. In the following we simply use the term `intermittent' to describe this particular subpopulation of AGN. 

So, after concluding that high accreting AGN are not primarily triggered by major mergers, we are turning the problem around and try now to constrain the number of merging inactive galaxies, which host an intermittent AGN. To quantify this aspect in terms of limits to be expected and to investigate dependencies, we express the fraction of merging inactive galaxies that currently host such an intermittent AGN $f_\mathrm{m,ina\,\&\,agn}$ via 

\begin{equation}
f_\mathrm{m,ina\,\&\,agn} = f_\mathrm{agn} \times f_\mathrm{m,agn} \times \frac{t_\mathrm{m}}{t_\mathrm{agn}} \text{.}
\end{equation}

Here, $f_\mathrm{agn}$ and $t_\mathrm{agn}$ describe the fraction and total duration of sources with accretion rates $>$70\% of the Eddington accretion rate during one single major merger event with regard to the overall galaxy population in the same redshift and mass bin. 
The merger fraction of the AGN subpopulation is given by $f_\mathrm{m,agn}$ and finally the parameter $t_\mathrm{m}$ describes the duration in which merger features are distinctly visible.

To constrain $f_\mathrm{agn}$, we use the logarithmic number densities of the total galaxy population and our high accreting AGNs at $z=2$. For our AGNs we adopt a representative value of $\Phi = -6.2$ Mpc$^{-3}$ mag$^{-1}$ by taking the mean I-band magnitude of our sample (see table~\ref{tab:qso_prop}) yielding $m_I = -27.0~\mathrm{mag}$, and using the results by \citet{13_ross_sdss_qso_lf}. To derive a comparable number density for the total galaxy population at $z=2$ and suitable for our mass range, we utilize the stellar mass function presented in \citet{15_henriques_mass_fct_evo}. We calculate the mean stellar host mass from our black hole mass estimates (see Sect.~\ref{subsec:Ina_sample}), which yields a logarithmic number density of $\Phi = -3.4$ Mpc$^{-3}$ mag$^{-1}$. Taking eventually the ratio of both number densities results in $f_\mathrm{agn} = 1.6 \times 10^{-3}$. To account for the spread in the black hole masses, we repeat our analysis with $f_\mathrm{agn} = 10^{-3}$ and $f_\mathrm{agn} = 2 \times 10^{-3}$, resulting from the minimum and maximum stellar host mass, respectively. 
It should be noted that we neither quantify the uncertainties introduced by using different stellar mass functions \citep[][and references therein]{16_conselice_diff_mass_fct}, nor the errors of the black hole mass determinations, which would result in an extended range of stellar host masses and number densities.

For $f_\mathrm{m,agn}$ we use our retrieved value of $f_\mathrm{m,agn}$=0.24 $\pm$ 0.09, but also reiterate our calculations for $f_\mathrm{m,agn}$ = [0.10, 0.40, 0.50, 0.60, 0.70]. We conduct our calculations for fixed values of $t_\mathrm{m}$ with $t_\mathrm{m} = 10^8$yr and $t_\mathrm{m} = 10^9$yr, with the latter setup resulting in an increase of $t_\mathrm{agn}$ by a factor of 10.
As for $f_\mathrm{agn}$, we relate $f_\mathrm{m,ina\,\&\,agn}$ with respect to the overall galaxy population in the same redshift range and stellar mass range. Considering that the merger fraction of inactive galaxies that we retrieve is $f_\mathrm{m,ina} \simeq 0.2$, this parameter has an upper bound of $f_\mathrm{m,ina\,\&\,agn} = 0.2$, which would imply that 100\% of the inactive galaxies that are currently undergoing a detectable major merger actually host an intermittent AGN. Conversely, $f_\mathrm{m,ina\,\&\,agn} = 0.0$ would be the lower bound and would correspond to no major merger triggers an AGN at any point in time. 

\begin{figure}[t]
\includegraphics[width = 8.5cm]{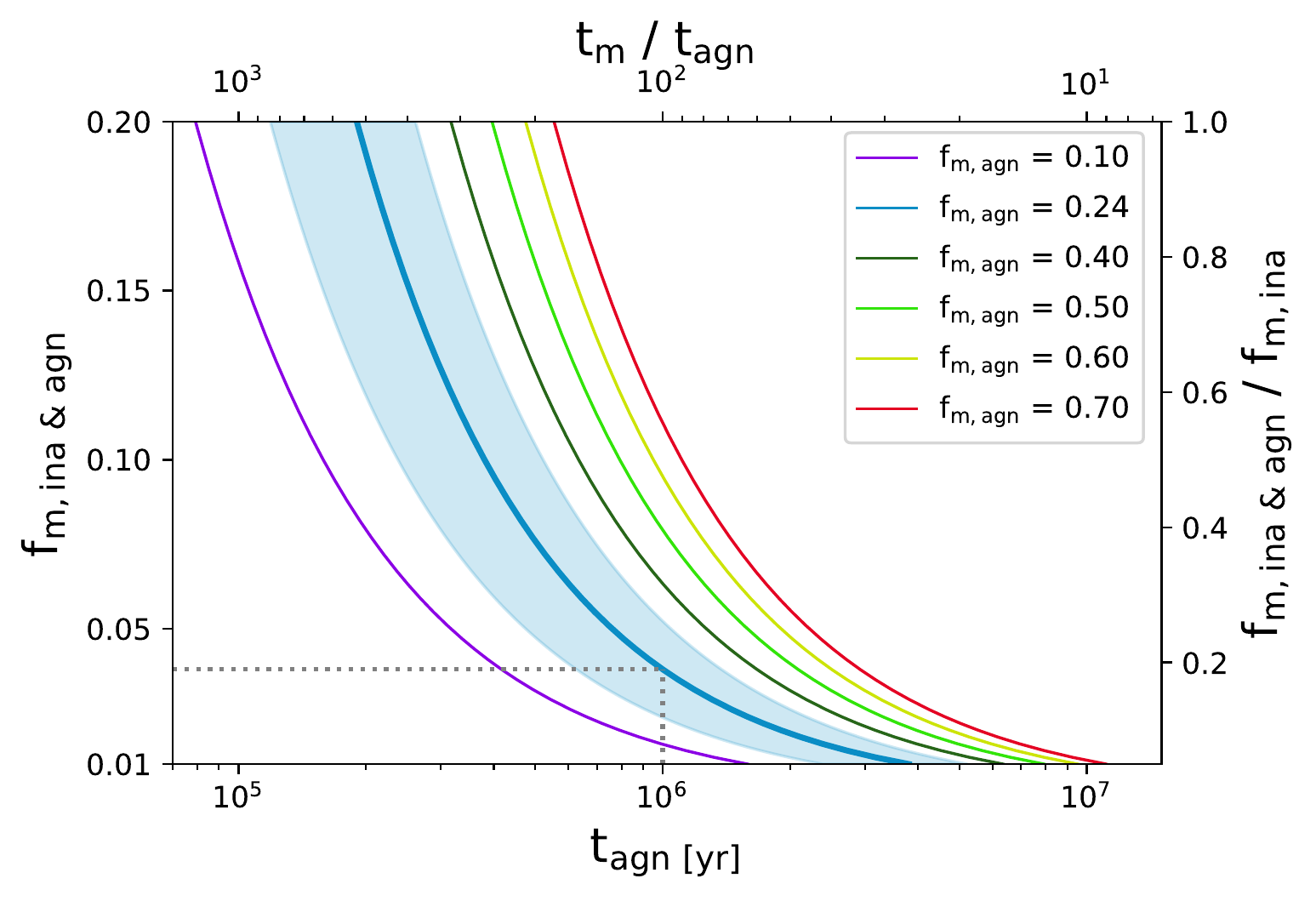}
\caption{Fraction of inactive galaxies with distinct merger features $f_\mathrm{m,ina\ \&\ agn}$ that host an intermittent AGN (see text for definition) versus the cumulative time $t_\mathrm{agn}$ that an AGN is accreting above 70\% of the Eddington accretion rate. On the secondary axes, we express the merger visibility time $t_\mathrm{m}$ as multiples of $t_\mathrm{agn}$, and $f_\mathrm{m,ina\ \&\ agn}$ normalized to the total fraction of inactive merging galaxies $f_\mathrm{m,ina}$. The blue line including the shaded region shows the results for our retrieved AGN merger fraction $f_\mathrm{m,agn}$ = 0.24 $\pm$ 0.09. The violet to red lines correspond to $f_\mathrm{m,agn}$ = [0.10, 0.40, 0.50, 0.60, 0.70]. 
\label{fig:hidden_agns}}
\end{figure}

In Fig.~\ref{fig:hidden_agns} we present the results of our calculation for different $f_\mathrm{m,agn}$ with $f_\mathrm{agn} = 1.6 \times 10^{-3}$ and $t_\mathrm{m} = 10^8$yr. On the X-axes we display the AGN timescale $t_\mathrm{agn}$ together with the merger timescale normalized over the AGN timescale $t_\mathrm{m}/t_\mathrm{agn}$. On the Y-axes we show the fraction of inactive, merging galaxies that host an intermittent AGN $f_\mathrm{m,ina\,\&\,agn}$, as well as this parameter normalized over the total subpopulation of inactive merging galaxies $f_\mathrm{m,ina\ \&\ agn} / f_\mathrm{m,ina}$. 
The blue line depicts our found AGN merger fraction, $f_\mathrm{m,agn} = 0.24$, whereas the blue shaded region displays the $1\sigma$ interval. The other colored values, ranging from violet to red correspond to the values for $f_\mathrm{m,agn}$ mentioned above. . 
A trend of increasing $f_\mathrm{m,ina\,\&\,agn}$ with decreasing $t_\mathrm{agn}$ is clearly visible. For a constant $t_\mathrm{agn}$, $f_\mathrm{m,ina\,\&\,agn}$ grows with increasing $f_\mathrm{m, agn}$. For our retrieved value of $f_\mathrm{m,agn}$ = 0.24, we obtain cumulative time scale limits in the range of $2\times10^5\lesssim t_\mathrm{agn} [\mathrm{yr}]\lesssim 4\times10^6$ for $f_\mathrm{m,ina\,\&\,agn} = 0.01$ and $f_\mathrm{m,ina\,\&\,agn} = 0.20$, respectively. 

If, in addition to $f_\mathrm{m,agn}$ = 0.24 $\pm$ 0.09 and $f_\mathrm{agn} = 1.6 \times 10^{-3}$, we now assume an rather short AGN timescale $t_\mathrm{agn} = 10^6$ yr, corresponding to 1\% of the merger time (or 0.1\% for a $10^9$ yr galaxy merger) those assumptions yield a fraction of inactive, merging galaxies that host an intermittent AGN of $f_\mathrm{m,ina\,\&\,agn} \sim 0.04 ^{{+0.01}}_{{-0.02}}$. This result implies that only $\sim 20\%$ of the comparison galaxies, that display distinct merger features are hosting an intermittent AGN, that is currently undetectable. For reference, we display in Fig.~\ref{fig:hidden_agns} our estimated $t_\mathrm{agn} = 10^6$ yr and the resulting $f_\mathrm{m,ina\,\&\,agn} \simeq 0.04$ by the dotted gray lines. 

Unfortunately, there are no detailed simulations available that model particular accretion rate histories, which could put a hard constrain on $t_\mathrm{agn}$. 
Hence, we resort to small-scale simulations following the growth of black holes in a more global context \citep{05_diMatteo_high_acc_duration, 09_johansson_high_acc_duration, 09b_johansson_high_acc_duration, 10_hopkins_high_acc_duration, 18_jung_high_acc_duration}. Our assumed cumulative time scale of $t_\mathrm{agn} = 10^6$ yr for a black hole that accretes higher than 70\% of the Eddington accretion rate, lies within the uncertainties provided by such simulations, but at the lower end of the expected value (see also Sect.~\ref{subsec:time_lag}). Assuming an AGN timescale an order of magnitude larger -- i.e.\ $t_\mathrm{agn} = 10^7$ yr -- already results in a fraction of inactive merging galaxies hosting an intermittent AGN that converges to zero, independent on the AGN merger fraction $f_\mathrm{m,agn}$. 

To be consistent with the theory suggested in \citet{18_goulding_high_lum_agn_conn_time_variability}, if all of the merging, inactive galaxies should host an AGN at some point (i.e.\ $f_\mathrm{m,ina\,\&\,agn} = 0.20$), the ratio of $t_\mathrm{agn}/t_\mathrm{m}$ would need to be smaller by a factor of 5-10, depending on $f_\mathrm{agn}$ and assuming our derived value for $f_\mathrm{m, agn}$ is correct. This would result in relatively short and therefore rather improbable AGN timescales of $1.2 \times 10^5 \lesssim t_\mathrm{agn} [\mathrm{yr}] \lesssim 2.4 \times 10^5 $. 
To put better directly physical constraints on those estimates, more detailed measurements and
especially simulations of merger timescales, accretion rate histories, and AGN populations
classified by their accretion rates are essential.

\subsection{Connection to post-starburst galaxies}

In recent years several studies found that post-starburst galaxies have small sizes and often show evidence for a recent merger in the form of tidal tails and distortions \citep{16_pawlik_merger_psb_galaxies, 16_wild_num_dens_ps_gal, 17_almaini_psb_galaxies_sizes}. By comparing number densities of such galaxies and our high accreting AGNs at $z=2$, and deriving the resulting AGN timescale, we want to examine if this particular population of galaxies is more prone to host AGNs and if the feedback of an active black hole could therefore be the dominant reason to quench the increased star formation. 

We adopt the same log number density for our AGNs as presented in Sect.~\ref{subsec:intermittence}, i.e.\ $\Phi \sim -6.2$ Mpc$^{-3}$ mag$^{-1}$ \citep{13_ross_sdss_qso_lf}.
For the post-starburst galaxies in our mass range at $z=2$ we use a corresponding number density of $\Phi \sim -4.2$ Mpc$^{-3}$ mag$^{-1}$ \citep{16_wild_num_dens_ps_gal}. The evident difference of two orders of magnitudes increases when we consider that not all AGNs can be hosted in post-starburst galaxies or their progenitors since we clearly detect AGNs hosted in non-distorted or disk-dominated systems, which results in an estimated overall difference in number density between our AGN and post-starburst galaxies by a factor of $2-4 \times 10^2$. This result combined with an assumed visibility of post-starburst signatures of $10^9 \mathrm{yr}$ \citep{16_wild_num_dens_ps_gal} relates to a timescale of \s 2.5-5 Myr in which the AGN accretes above our threshold of 70\% Eddington ratio. As described already in the previous two sections such a short timescale appears improbable leading us to the conclusion that even in post-starburst galaxies, where AGNs could play understandably a prominent role due to the obvious merger processes and the need to quench the elevated star formation, active black holes still play only a minor role. 
Even if we consider lower accretion rates of approximately a few percent of the Eddington ratio,
and hence lower luminosities, the corresponding number density and resulting AGN timescale would only increase by a factor of \s 3, implying still a very improbable lifespan for AGNs with such or higher accretion rates.

\subsection{Alternative mechanisms}

We conclude that major mergers are not only a sub-dominant driver for the triggering of AGNs
but also, that the majority of major mergers do not experience an AGN.
By ruling out major mergers as the prevailing mechanism to produce even AGN with the highest Eddington ratios, the question arises which mechanisms could dominate triggering and fueling of AGNs.
Viable options are minor mergers/satellite infall, and/or secular, internal instabilities \citep{09_georgakakis_AGN_disc_gal_not_mergers, 11_allevato_x_ray_bl_agn_intermediate_lum_no_merger, 11_bournoaud_disk_instability, 11_cisternas_BH_Growth_z_1, 12_kocevski_merger_connection_z=2_x_ray, 17_villforth_host_gal_lum_z_0.6_no_merger_con}. Although initially presuming a redshift-independent bimodality in the form that minor mergers are mainly responsible for AGN with low luminosities \citep{06_hopkins_minor_merger, 09b_hopkins_minor_merger} while major mergers dominate the growth of luminous AGNs \citep{14_hopkins_redshift_lum_agn_host_morphology}, there have been recent studies suggesting that minor mergers could be connected to luminous AGN after all \citep{09_georgakakis_AGN_disc_gal_not_mergers, 11_allevato_x_ray_bl_agn_intermediate_lum_no_merger, 17_villforth_host_gal_lum_z_0.6_no_merger_con}. 

Recent studies show no correlation between increased AGN activity and central bars, which could also lead to disk-instabilities and subsequently to a gas inflow to the galaxy center \citep{15_cheung_agn_bar_connection_mod_lum, 15_cisternas_agn_bar_connection, 17_goulding_agn_bar_connection}. This leaves minor mergers and bar-independent secular processes as the remaining viable options for the bulk of black hole growth.
However, due to the faint signatures of minor mergers, it has not yet been possible to verify this theory to a statistical significant extent. 
With the launch of the James Webb Space Telescope (JWST) this will change and it will be possible to detect not only features arising from major mergers at larger redshifts, but also from minor mergers. A study designed similar to our analysis can then give insights if minor mergers are more frequent in galaxies hosting a luminous AGN, answering if such processes are predominant in the triggering of all AGN.

\section{Summary \& Conclusions} \label{sec:summary}

In this first study specifically targeting black holes with the highest specific accretion rates, we test the merger--AGN connection at $z\sim2$ by examining and comparing 21 AGN host galaxies and 92 matched inactive comparison galaxies. We simultaneously model a point-source component and a S\e rsic profile to infer the best fit for the AGN component as embedded in the galaxy. To create a sample of inactive galaxies with comparable residuals we add synthetic point sources at the respective flux centers and repeat our fitting process. As a result, we have a joint sample of 113 host galaxies, which were ranked from most to least distorted by 10 experts. We created three consensus rankings by combining those individual rankings with three different methods. Subsequently, we divided them into 
galaxies displaying features of a recent or ongoing merger event, and another subset showing no distinct features. In conclusion, we compare the merger fractions of AGN hosts and control galaxies.

In summary we find:
\begin{itemize}
\renewcommand\labelitemi{--}
\item We detect no significant enhancement of merger events for the AGN host galaxies with respect to a sample of matched inactive galaxies. The corresponding merger fractions are $f_\mathrm{m,agn}$ = 0.24 $\pm$ 0.09 and $f_\mathrm{m, ina} = $0.19 $\pm$ 0.04 for the AGN sample and inactive galaxies, respectively.
\item Neither the choice of cut-off rank nor the method used to combine the individual rankings has an impact on the derived merger fractions
\item Within our selection, the merger fractions do not strongly depend on stellar mass for either sample.
\item The S\e rsic indices of the AGN host galaxies and inactive galaxies show a negligible difference. We find a comparable fraction of disk-dominated galaxies between the two samples.
\item Considering all these observations in combination with results from other studies, we conclude that even for AGN at $z\sim2$ with the highest specific accretion rates, major mergers are not the dominant trigger. 
\item Taking into account both the timescale of AGN activity and merger visibility, we show that it is improbable that the lack of merger enhancement in our study is due to a time lag, that washes out merger features before the AGN phase ends.
\item We conclude that only a minority of AGN are triggered by major mergers, and need better simulated accretion-rate histories to constrain the numbers of merging galaxies producing an AGN. We consider intermittent AGN, which had and will again show high accretion rates, but are or at least appear currently inactive. 
With our estimates only $\sim$20\% within the population of merging and inactive galaxies would actually be hosting such an intermittent AGN. 
\item Considering the respective number densities and resulting AGN timescales, even post-starburst galaxies showing significant merger features and increased star formation rates appear not to be predominantly connected to AGNs.
\end{itemize}

Our results are consistent with the findings of previous studies, showing that independent of the AGN properties (luminosity, black hole mass, accretion rate), alternative mechanisms are needed to trigger and fuel the growth of black holes. Once active, black holes can have a significant impact on the subsequent evolution of its host galaxy, however, how AGNs are actually triggered to became active, still remains inconclusive. 

With observations of AGN host galaxies at $z\sim2$, we pushed HST to its limit in terms of resolution and available wavelength range. Besides continuing our analysis for galaxies hosting high-accretion BH at $z\sim0.2$, we will make use of the extraordinary capabilities of JWST in the future. We plan to do a more detailed analysis by examining more and also fainter galaxies, which will also give us the opportunity to detect weaker merger features -- up to signs caused by minor mergers -- to finally solve the open questions concerning how AGNs are triggered and fueled.

\acknowledgments

VM acknowledges support by the German Aerospace Center (DLR) under grant 50\,OR\,1609. RAW acknowledges NASA JWST grants NAG5-12460, NNX14AN10G and
80NSSC18K0200 from GSFC.

Based on observations made with the NASA/ESA Hubble Space Telescope, obtained from the Data Archive at the Space Telescope Science Institute, which is operated by the Association of Universities for Research in Astronomy, Inc., under NASA contract NAS 5-26555. Support for program number 14262 was provided by NASA through a grant from the Space Telescope Science Institute, which is operated by the Association of Universities for Research in Astronomy, Inc., under NASA contract NAS5-26555.

Funding for the Sloan Digital Sky Survey (SDSS) has been provided by the Alfred P. Sloan Foundation, the Participating Institutions, the National Aeronautics and Space Administration, the National Science Foundation, the U.S. Department of Energy, the Japanese Monbukagakusho, and the Max Planck Society. The SDSS Web site is http://www.sdss.org/.

The SDSS is managed by the Astrophysical Research Consortium (ARC) for the Participating Institutions. The Participating Institutions are The University of Chicago, Fermilab, the Institute for Advanced Study, the Japan Participation Group, The Johns Hopkins University, Los Alamos National Laboratory, the Max-Planck-Institute for Astronomy (MPIA), the Max-Planck-Institute for Astrophysics (MPA), New Mexico State University, University of Pittsburgh, Princeton University, the United States Naval Observatory, and the University of Washington. 

This work is based on observations taken for the dedicated GO program 14262 and by the 3D-HST Treasury Program (GO 12177 and 12328) with the NASA/ESA HST, which is operated by the Association of Universities for Research in Astronomy, Inc., under NASA contract NAS5-26555.

This research has made use of NASA's Astrophysics Data System Bibliographic Services

\facility{HST(WFC3)}
\software{
        astropy \citep{2013_astropy_I,2018_astropy_II},
        Matplotlib \citep{matplotlib},
        SAOImageDS9 \citep{saods9}
}

\newpage

\appendix

\section{AGN host galaxy modeling}\label{appendix:qso_models_figs}

The additional 20 AGNs of our sample, as in Fig.~\ref{fig:qso_showcase}. From left to right we show the original, processed HST/WFC3 image, the PSF convolved point-source model, and the host galaxies after point-source subtraction with the original sampling and smoothed by a $\sigma$=2px Gaussian, respectively. All images are displayed with the same arcsinh scaling.

\begin{figure*}[b]
\centering
\vspace{1.5ex}
\includegraphics[width = 18cm]{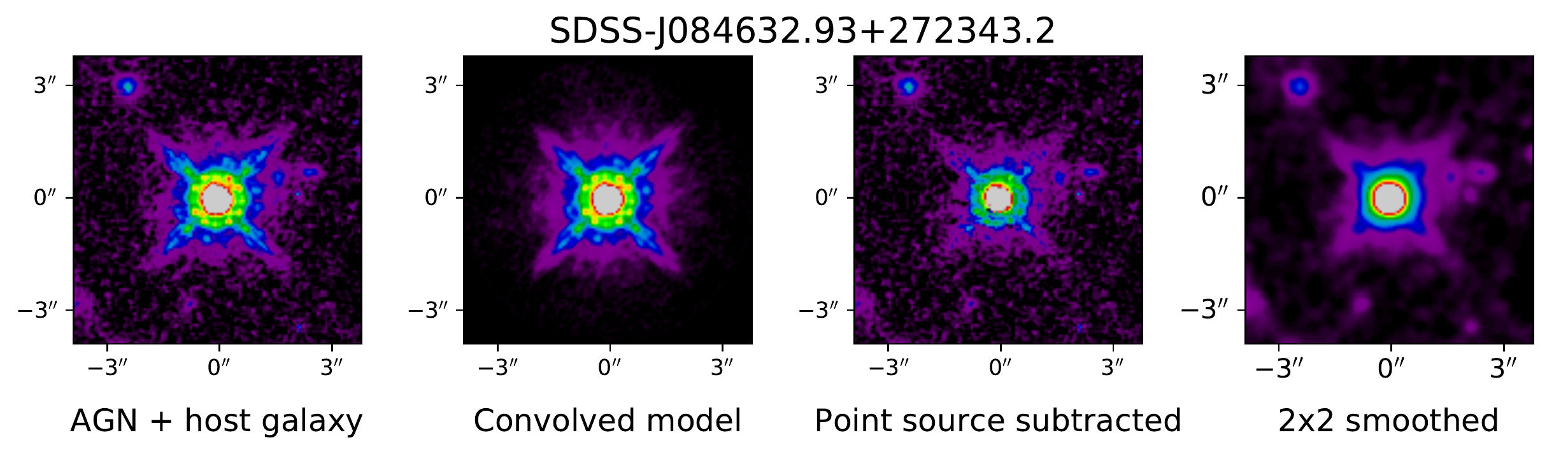}
\label{fig:qso_1}
\end{figure*}

\begin{figure*}[b]
\centering
\includegraphics[width = 18cm]{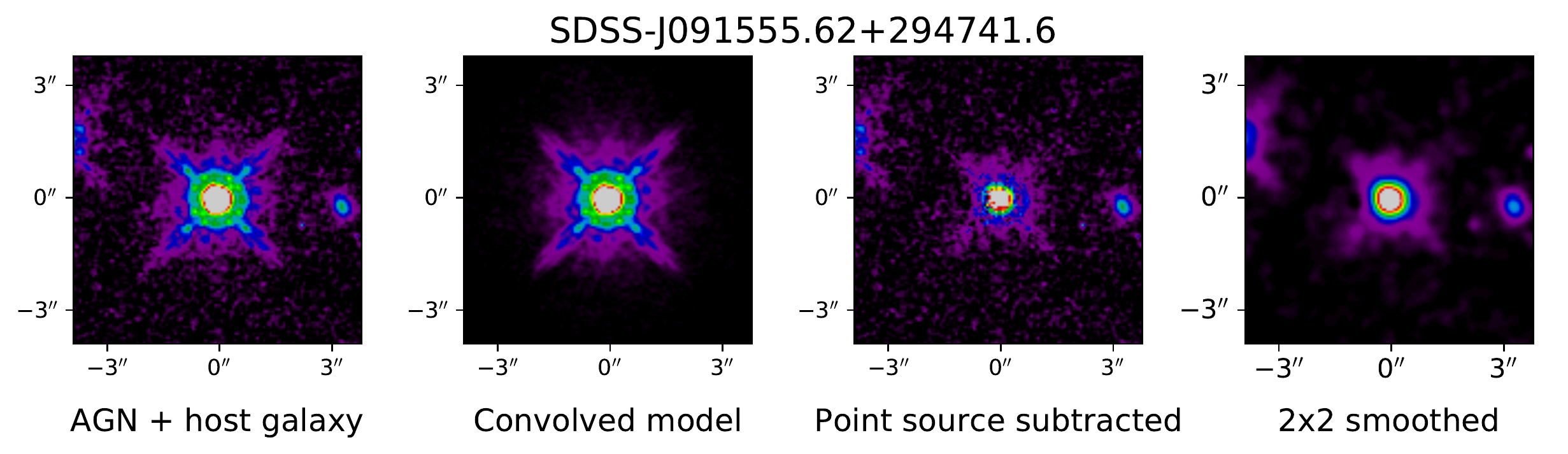}
\end{figure*}

\begin{figure*}[b]
\centering
\includegraphics[width = 18cm]{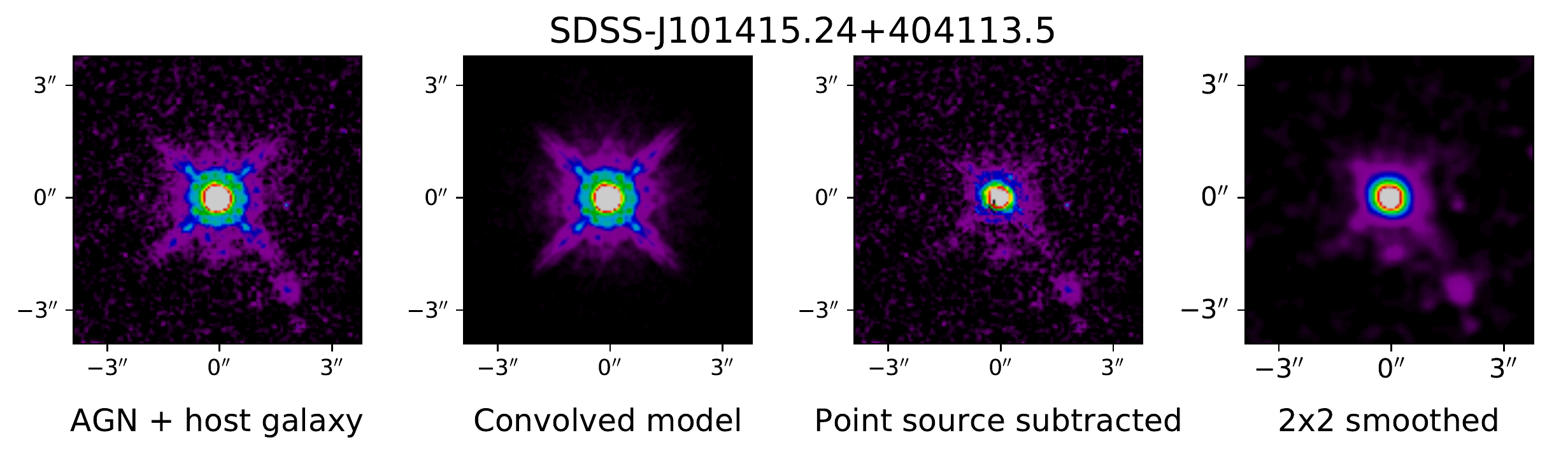}
\end{figure*}

\begin{figure*}[b]
\centering
\includegraphics[width = 18cm]{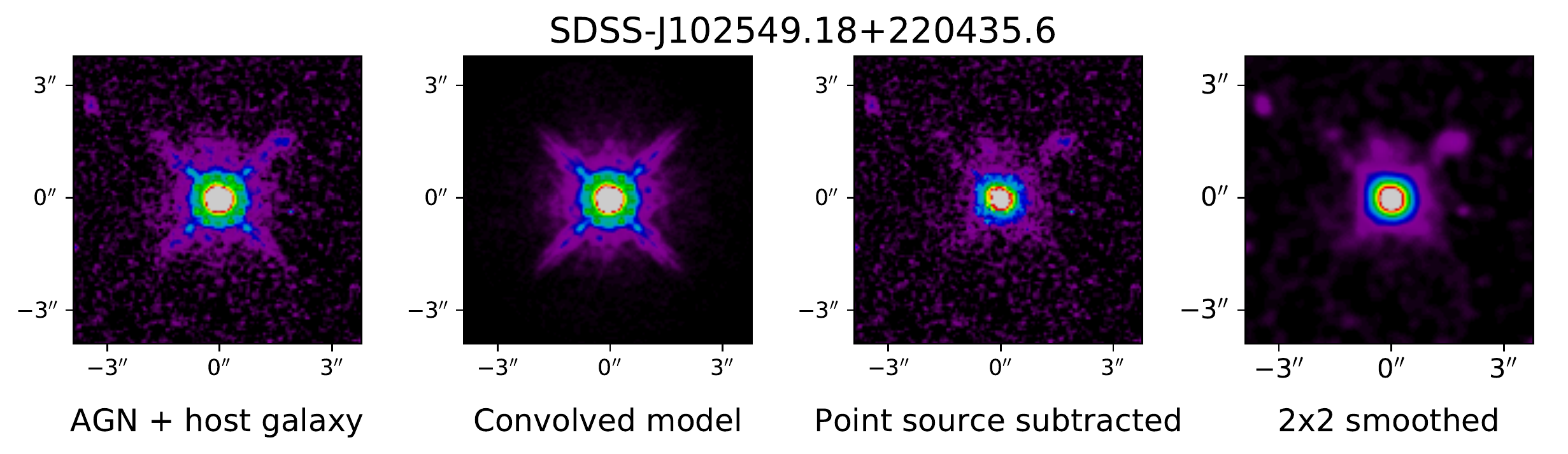}
\end{figure*}

\begin{figure*}[b]
\centering
\includegraphics[width = 18cm]{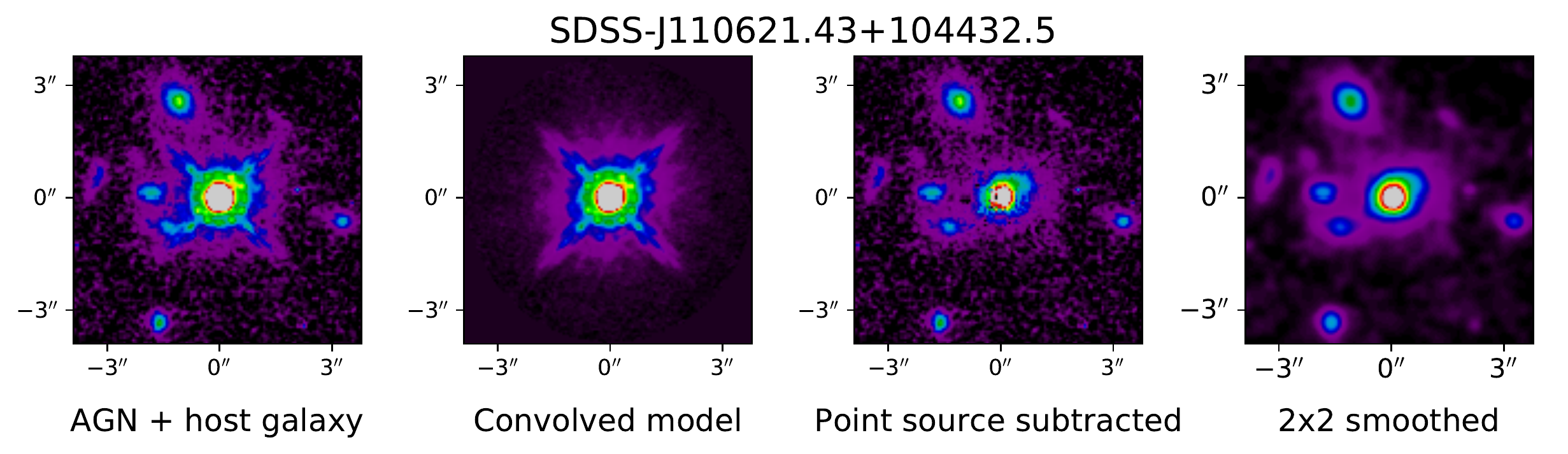}
\end{figure*}

\begin{figure*}[b]
\centering
\includegraphics[width = 18cm]{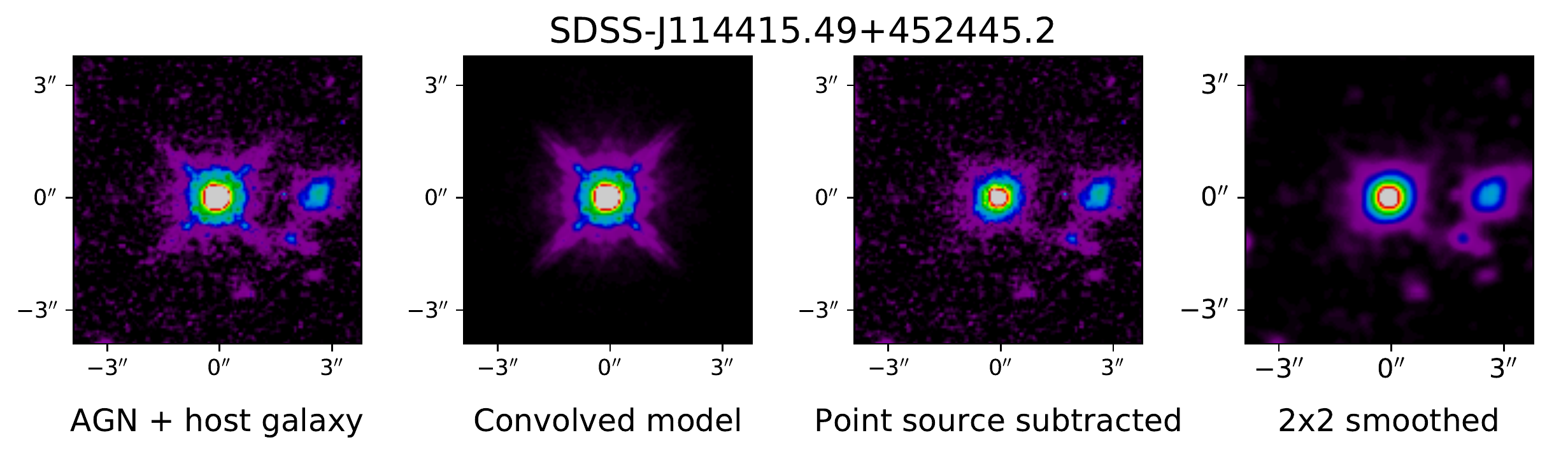}
\end{figure*}

\begin{figure*}[b]
\centering
\includegraphics[width = 18cm]{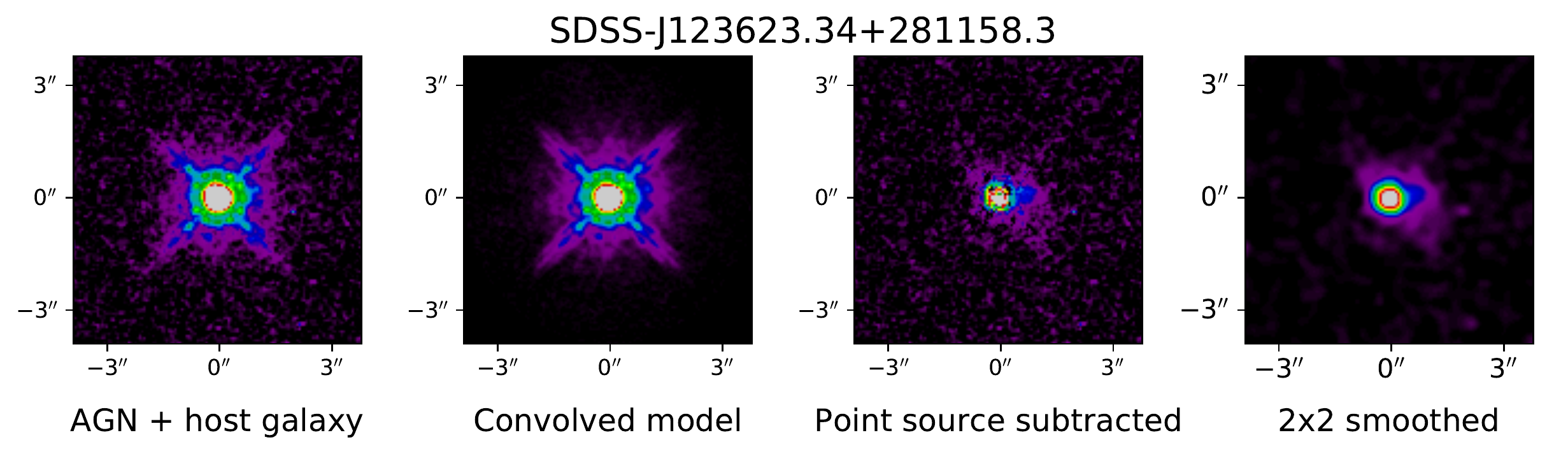}
\end{figure*}

\begin{figure*}[b]
\centering
\includegraphics[width = 18cm]{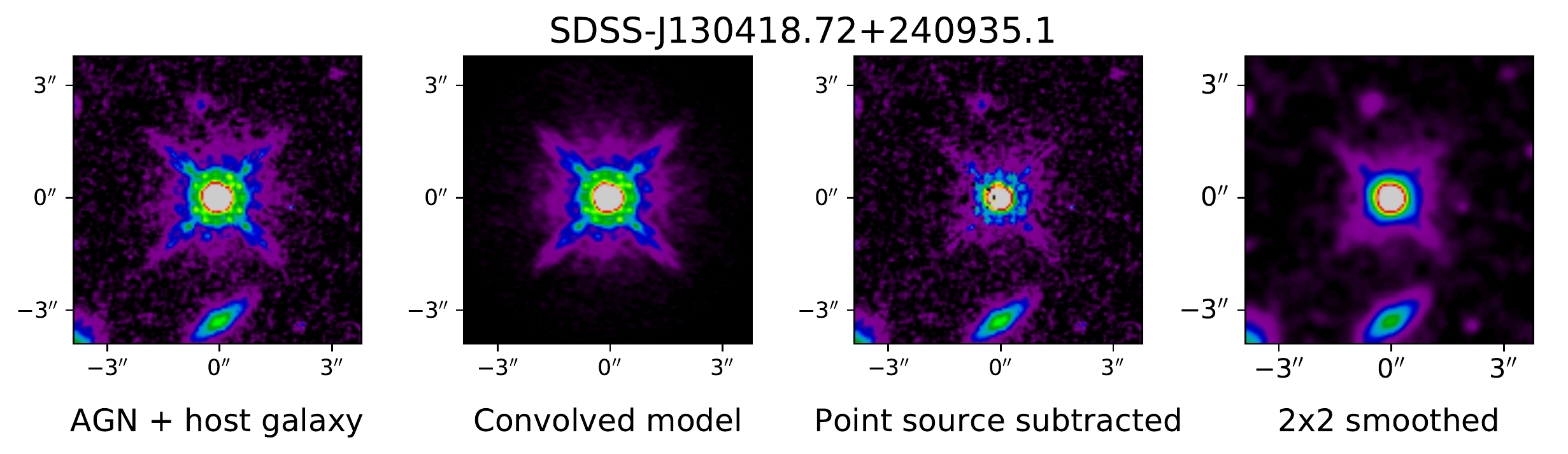}
\end{figure*}

\begin{figure*}[b]
\centering
\includegraphics[width = 18cm]{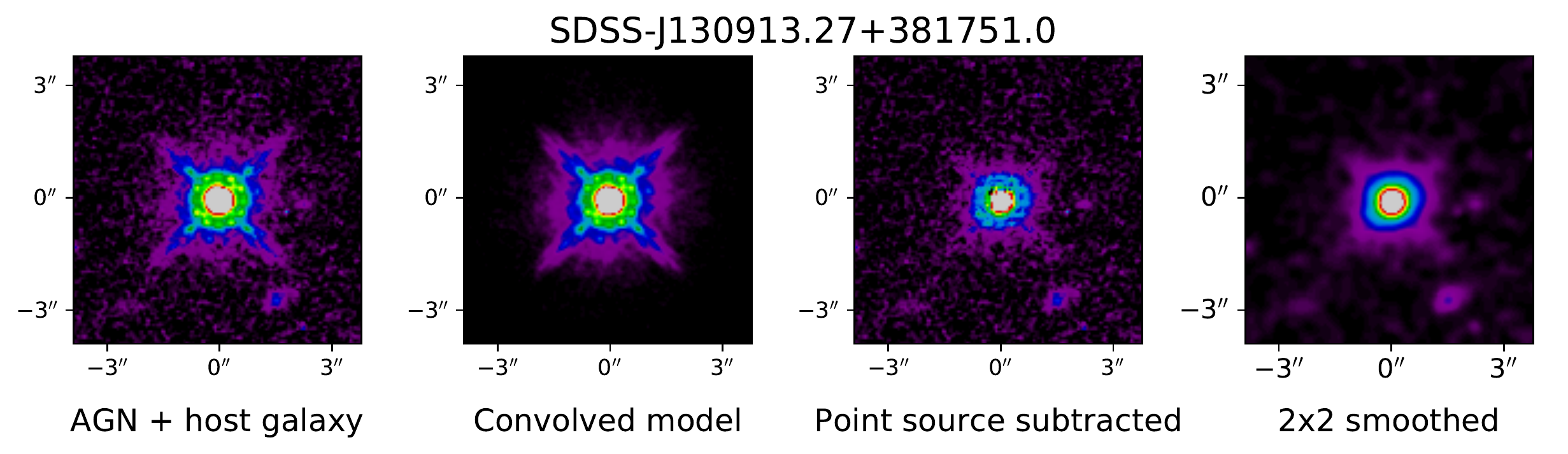}
\end{figure*}

\begin{figure*}[b]
\centering
\includegraphics[width = 18cm]{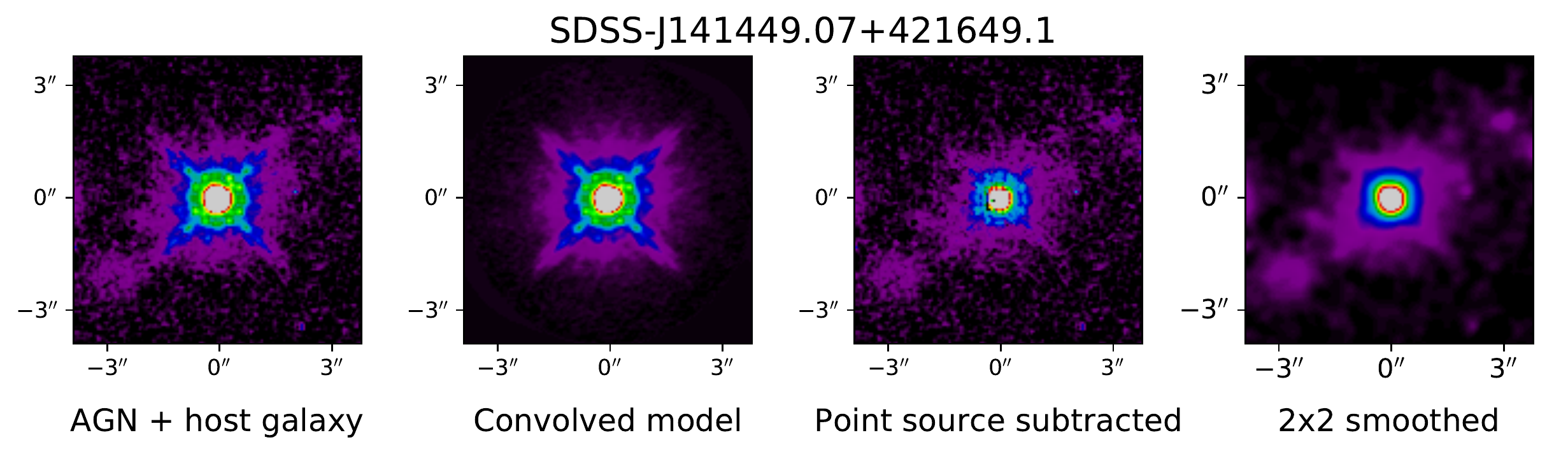}
\end{figure*}

\begin{figure*}[b]
\centering
\includegraphics[width = 18cm]{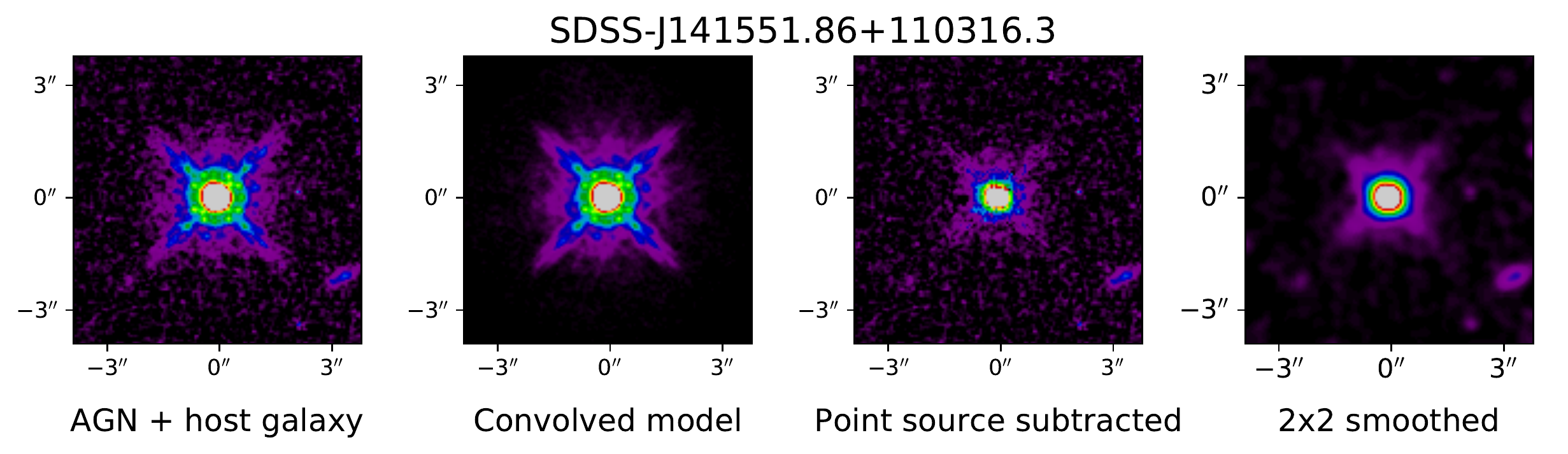}
\end{figure*}

\begin{figure*}[b]
\centering
\includegraphics[width = 18cm]{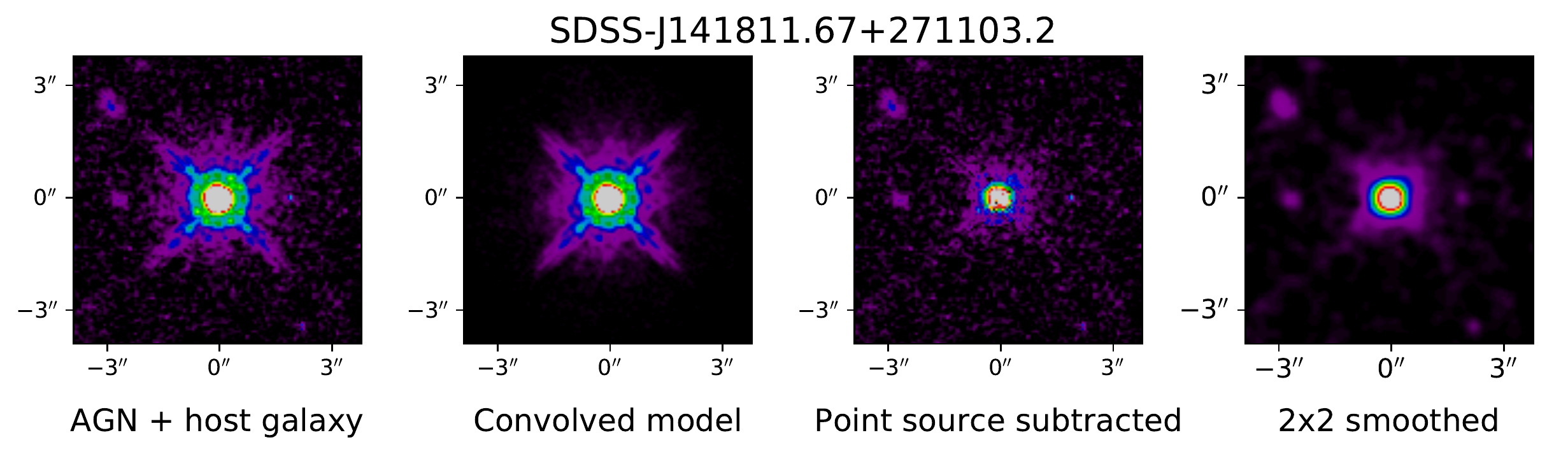}
\end{figure*}

\begin{figure*}[b]
\centering
\includegraphics[width = 18cm]{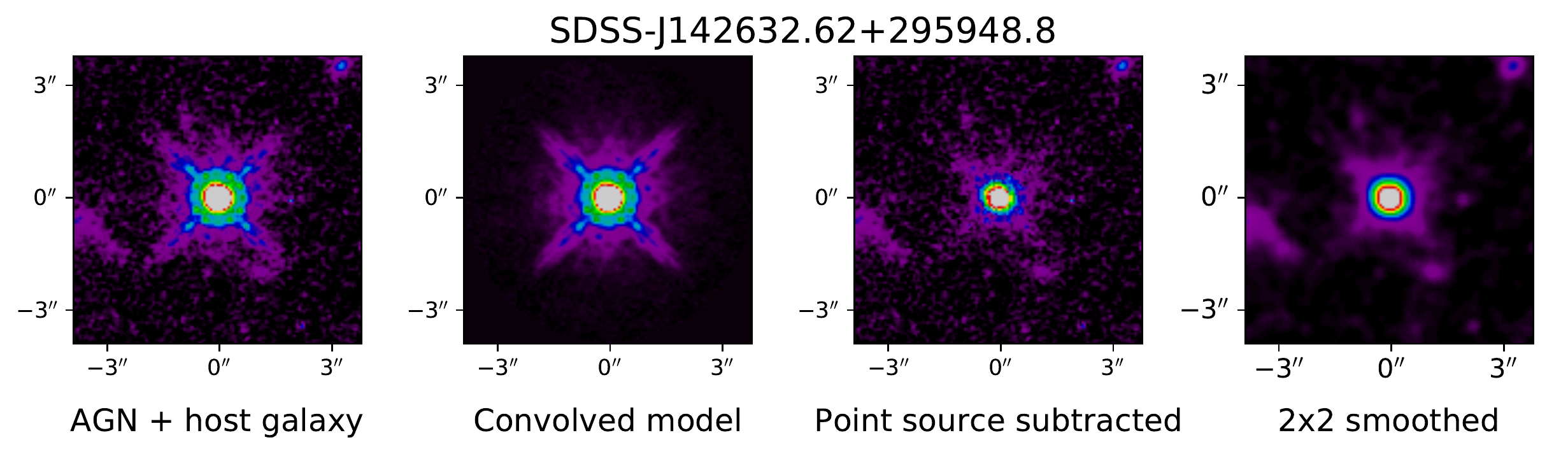}
\end{figure*}

\begin{figure*}[b]
\centering
\includegraphics[width = 18cm]{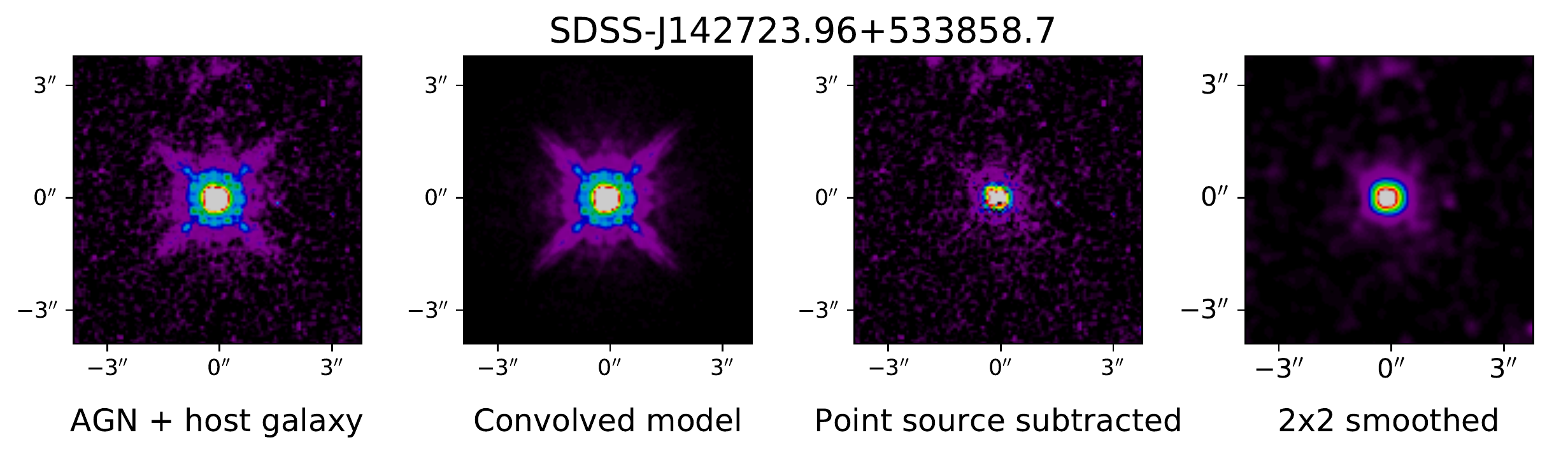}
\end{figure*}

\begin{figure*}[b]
\centering
\includegraphics[width = 18cm]{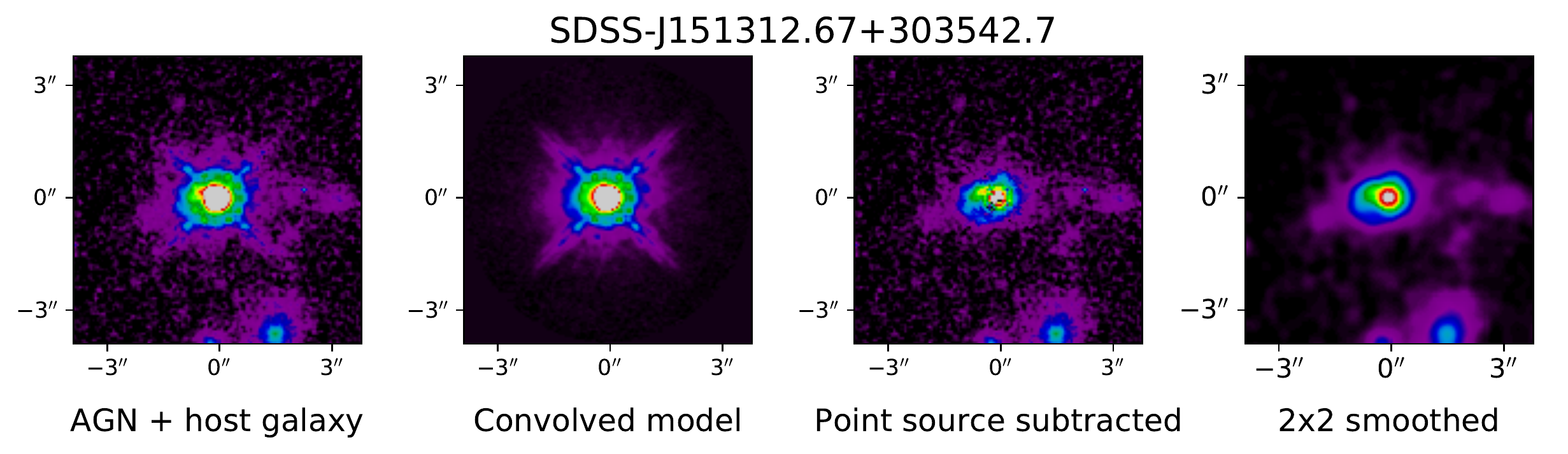}
\end{figure*}

\begin{figure*}[b]
\centering
\includegraphics[width = 18cm]{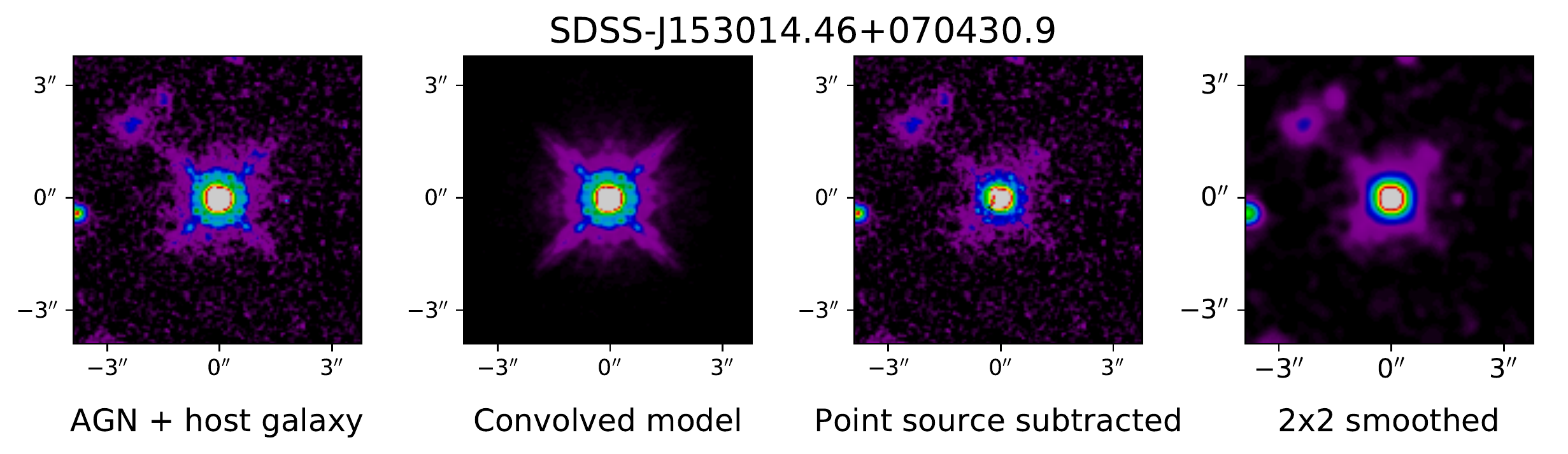}
\end{figure*}

\begin{figure*}[b]
\centering
\includegraphics[width = 18cm]{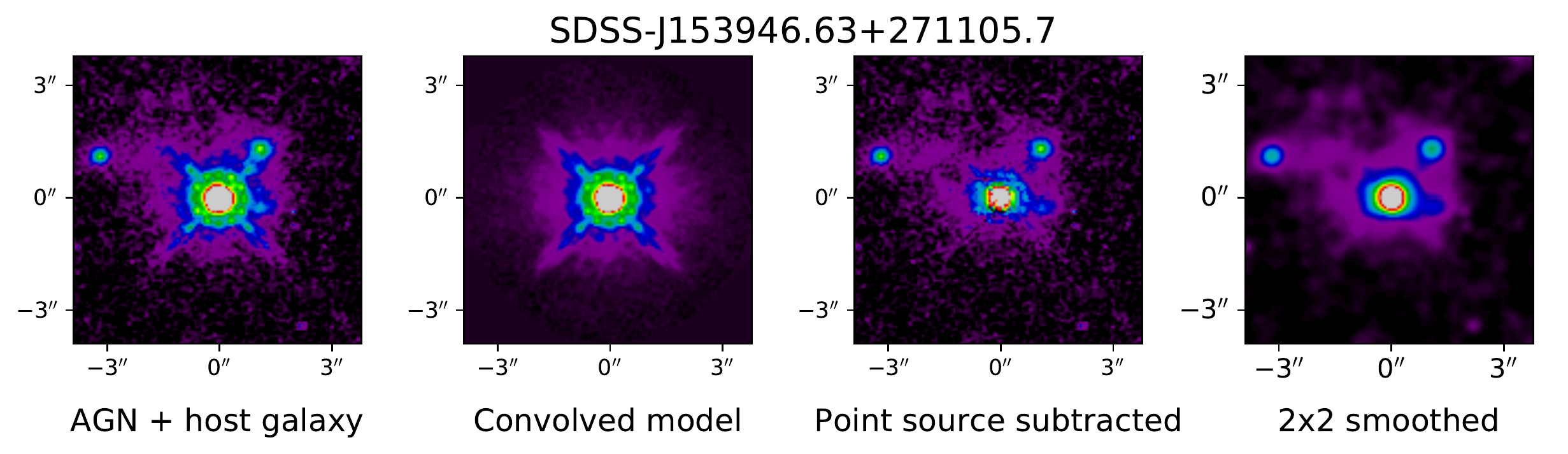}
\end{figure*}

\begin{figure*}[b]
\centering
\includegraphics[width = 18cm]{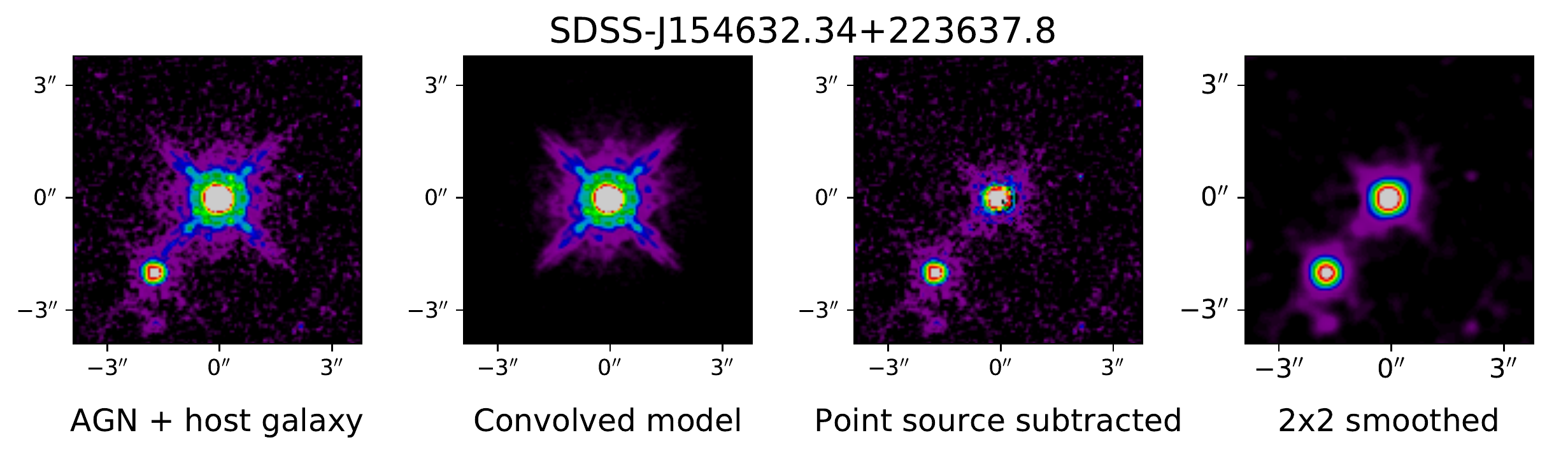}
\end{figure*}

\begin{figure*}[b]
\centering
\includegraphics[width = 18cm]{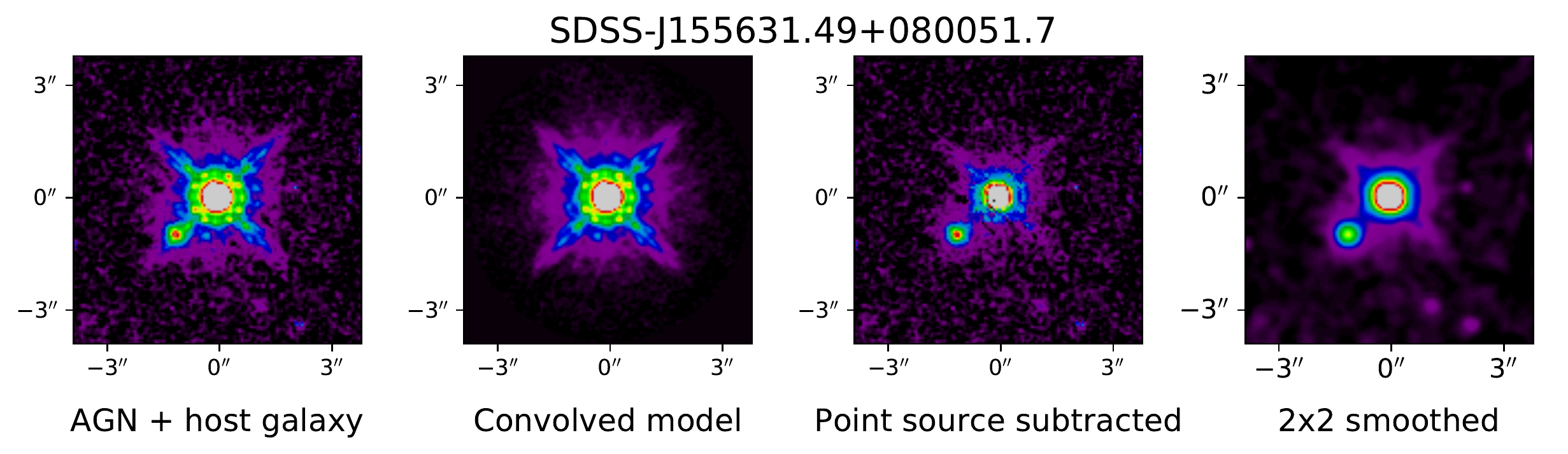}
\end{figure*}

\begin{figure*}[t]
\centering
\includegraphics[width = 18cm]{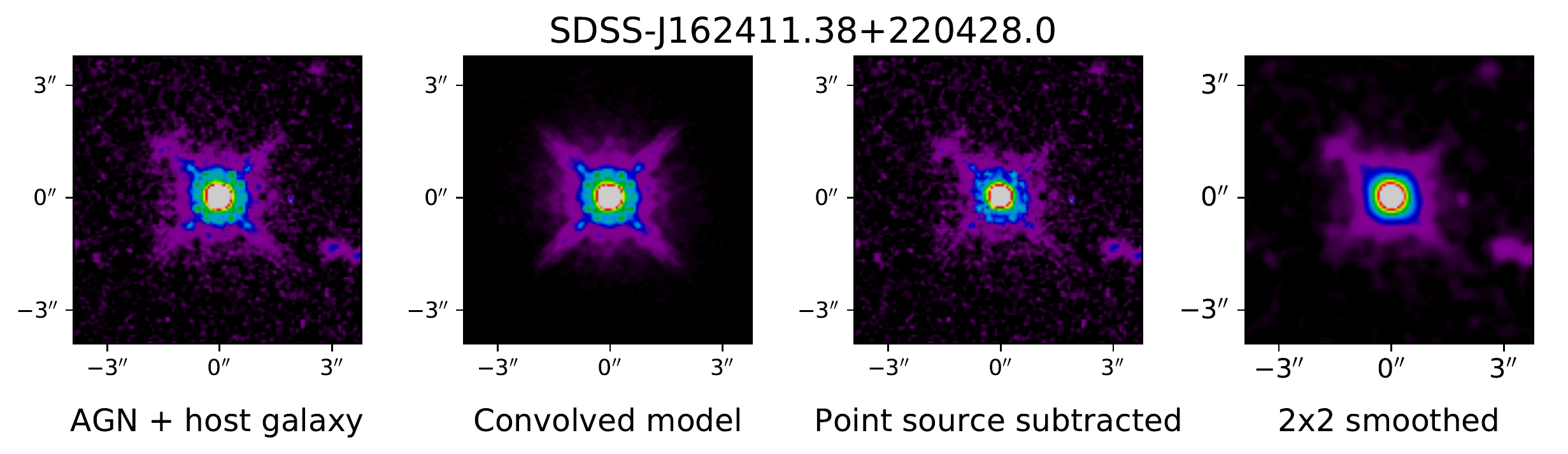}
\end{figure*}

\newpage

\section{Details on combination methods}\label{appendix:details_methods}

As our first method to combine the individual expert rankings, we choose the same approach as in \citet{16_mechtley_agn_merger_most_massive}. We calculate the mean rank for each galaxy, and discard the 23 individual assessments that deviate by more than 2$\sigma$ from the respective average ranks. However, it should be noted that this method violates the non-dictatorship condition, one of a set of criteria summarized in Arrow's Impossibility Theorem \citep{50_arrow_impossibility_theorem}. It states that no method exists for at least two voters having three or more options that satisfies the following three axioms: (1) non-dictatorship, such that all individual votes are weighed equally; (2) the weak pareto principle or unanimity, such that if all voters elect $x > y$ this condition is also met in the overall ranking; and (3) independence of irrelevant alternatives, such that the consensus preference between $x$ and $y$ only depends on the individual preferences between $x$ and $y$ and not on (an) additional option(s). 

Therefore we choose the Borda \citep{13_emerson_borda_count} count, part of electoral systems, as our second method. It satisfies the non-dictatorship, but violates in turn the independence of irrelevant alternatives condition. In the original version, the overall ranking is created by appointing every candidate points according to its individual rank. The first ranked option receives $n$ points, the second preference $n-1$ and so on, with $n$ being the number of candidates. To decrease the impact of low ranked galaxies, as they may be ranked more randomly due to a lack of merger features, we applied a variant of this system called the Dowdall System \citep{02_reilly_dowdall_sys}. By calculating the respective points using the reciprocal value of the respective ranks, i.e.\ $1/n$, a top-ranked galaxy gets one point, the next one 0.5 points and so on, while flagged entries were given 0 points. This method avoids the so-called Condorcet's paradox \citep{1785_condorcet_method, 89_condorcet_translation}, a case in which a consensus ranking can be cyclic, although the individual preferences are not. Still, it does not satisfy the Condorcet criterion, which states that the overall top-ranked candidate wins in every pairwise comparison of candidates. 

Hence, we apply the Schulze method \citep{11_schulze_schulze_method, 18_schulze_schulze_method} as our third approach a system, which satisfies this criterion. A comparison of the pairwise preferences, i.e.\ how many voters prefer $x$ over $y$ gives a quantitative evaluation for each candidate, resulting in a sequence where a candidate wins over all other candidates, a so-called Condorcet winner, whereas the overall second placed option only loses to the winner and so on \citep[for more details on the method please see][]{18_schulze_schulze_method}. We treat individual flagged entries as unranked votes and make no preferences within them.

\section{Assessing potential influences by choice of combination method or cut-off rank}\label{appendix:fraction_dependences}

\begin{figure*}[t]
\centering
\includegraphics[width = 18cm]{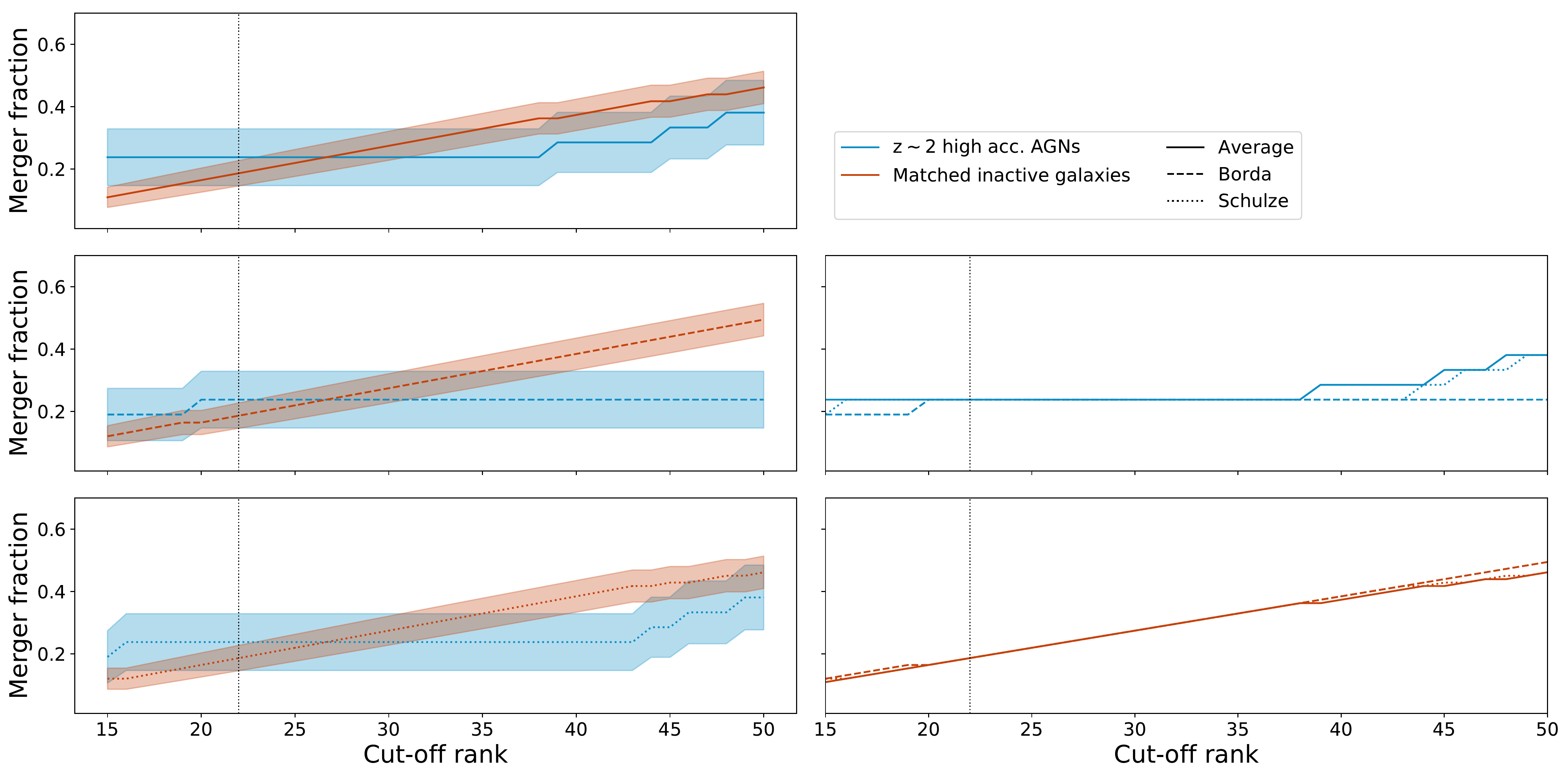}
\caption{Distributions of the merger fractions of AGN host galaxies and inactive galaxies versus of cut-off rank and method to derive the consensus ranking. The blue and red color represent the AGN and comparison sample respectively. The line-style denotes the respective combination method. The shaded regions give the $1\sigma$ interval. The plots on the left display the differences in merger fraction for both samples, depending on the applied method. From top to bottom we show the Average, Borda count and Schulze method. The two plots on the right show the three methods combined for the two respective samples (AGN sample in blue in the upper, comparison sample in red in the lower plot).
\label{fig:cut_off_rank_depedency}}
\end{figure*}

To examine any influence resulting from the applied method to combine the individual rankings or the choice of cut-off rank, we recalculate the merger fractions for a sequence of cut-off ranks between 15 and 50 and assess the resulting values for each of the three methods. Our assessments are visualized in Fig.~\ref{fig:cut_off_rank_depedency}, where the gray dotted vertical line denotes our chosen cut-off rank at position 22, the color scheme describes the respective sample, blue for AGN and red for the comparison sample, whereas the line type depicts the respective method: solid for the average method, dashed for the Borda count approach and dotted for the Schulze method. Hence, the three plots on the left hand side show the evolution of the merger fractions for both samples versus the chosen cut-off rank for each method. The two diagrams on the right display the results on the respective merger fractions for each method. By definition the fractions -- especially for the inactive sample -- increase the higher the chosen cut-off rank is. However, except for the Borda count at cut-off ranks between 40 and 50, the merger fractions for both samples are less than 1$\sigma$ different for each individual method.

While it appears that there is a slight, but not significant lack of enhancement for the AGN merger fraction derived with the Borda method with respect to the other two methods, the fractions for the inactive galaxies are practically identical.

The identical results for the three combination methods between ranks 20 and 40 underline two facts: First, there is a very good overall agreement within the individual rankings regarding which one is a merging galaxy and which is a non-disturbed system. Otherwise, we would see slight discrepant results like we got for the ranks $\geqslant$ 40 for the AGN merger fractions. These slight deviations however are simply explained by the fact that every expert ranks the inconspicuous galaxies marginally different, since they show no merger features whatsoever, and that the distinct algorithms of the three combination methods yield slightly varying results. Secondly, this already implies that -- even without visually examining the objects -- the cut-off rank must lie between the ranks 20 and 40. Hence, we chose our cut-off rank well, and it lies in a range where the merger fractions are indeed identical, confirming that the choice of combination method has no impact on the final results. 
Furthermore, it should also be noted that a cut-off rank of 50 corresponds to a merger fraction of \s 50\%, a value, which has not been detected by any observation or simulation.  

\end{document}